\documentclass[twoside,twocolumn,9pt]{article}

\usepackage[super,sort&compress,comma]{natbib} 
\usepackage[version=3]{mhchem}
\usepackage[left=1.5cm, right=1.5cm, top=1.785cm, bottom=2.0cm]{geometry}
\usepackage{balance}
\usepackage{mathptmx}
\usepackage{sectsty}
\usepackage{graphicx} 
\usepackage{lastpage}
\usepackage[format=plain,justification=justified,singlelinecheck=false,font={stretch=1.125,small,sf},labelfont=bf,labelsep=space]{caption}
\usepackage{float}
\usepackage{fancyhdr}
\usepackage{fnpos}
\usepackage[english]{babel}
\addto{\captionsenglish}{%
  
}

\DeclareFontFamily{OT1}{pzc}{}
\DeclareFontShape{OT1}{pzc}{m}{it}{<-> s * [1.10] pzcmi7t}{}
\DeclareMathAlphabet{\mathpzc}{OT1}{pzc}{m}{it}
\usepackage{array}
\usepackage{droidsans}
\usepackage{charter}
\usepackage[T1]{fontenc}
\usepackage[usenames,dvipsnames]{xcolor}
\usepackage{setspace}
\usepackage[compact]{titlesec}
\usepackage{hyperref}
\usepackage{xcolor}

\usepackage{epstopdf}

\definecolor{cream}{RGB}{222,217,201}

\usepackage{tikz,xcolor,hyperref}

\definecolor{lime}{HTML}{A6CE39}

\newcommand{\orcid}[1]{\href{https://orcid.org/#1}{\textcolor[HTML]{A6CE39}{\aiOrcid}}}

\begin{document}
\setstretch{1.125} 
\makeatletter 
\newlength{\figrulesep} 
\setlength{\figrulesep}{0.5\textfloatsep} 
\newcommand{\topfigrule}{\vspace*{-1pt}%
\noindent{\color{cream}\rule[-\figrulesep]{\columnwidth}{1.5pt}} }
\newcommand{\botfigrule}{\vspace*{-2pt}%
\noindent{\color{cream}\rule[\figrulesep]{\columnwidth}{1.5pt}} }
\newcommand{\dblfigrule}{\vspace*{-1pt}%
\noindent{\color{cream}\rule[-\figrulesep]{\textwidth}{1.5pt}} }
\makeatother

\twocolumn[
  \begin{@twocolumnfalse}
  \LARGE{\textbf{Pressure-deformation relations of elasto-capillary drops (droploons) on capillaries}} \vspace{0.6cm} \\
  
  \noindent\large{Ga\"el Ginot\textit{$^{a}$}}\href{https://orcid.org/0000-0002-9006-0319}{\includegraphics[width=0.3cm]{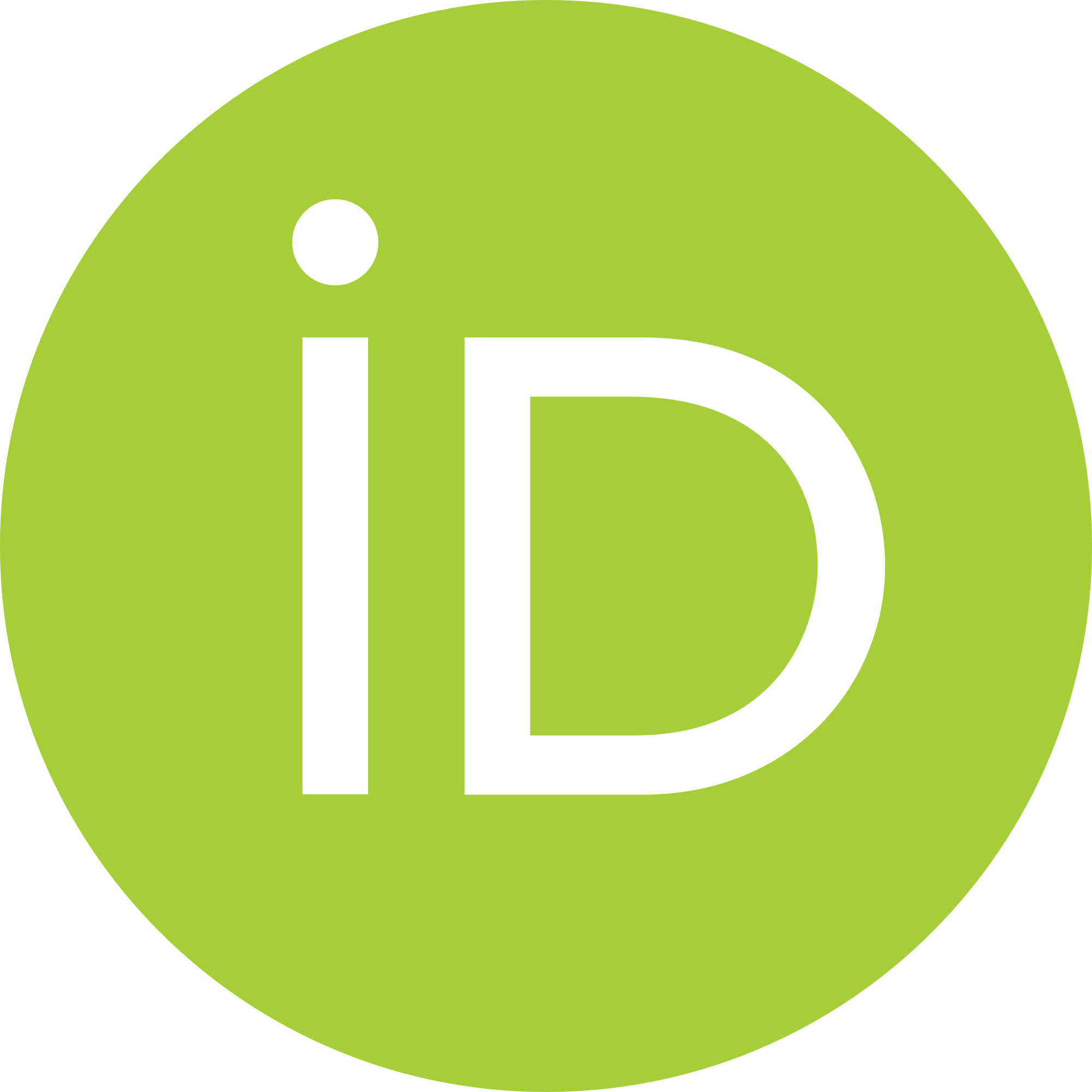}},  Felix S. Kratz\textit{$^{d}$}\href{https://orcid.org/0000-0001-9509-1299}{\includegraphics[width=0.3cm]{orcid_logo.png}}, Friedrich Walzel\textit{$^{a}$}, Jean Farago\textit{$^{a}$},  Jan Kierfeld\textit{$^{d}$}\href{https://orcid.org/0000-0003-4291-0638}{\includegraphics[width=0.3cm]{orcid_logo.png}}, Reinhard H\"ohler$^{\ast}$\textit{$^{b,c}$}\href{https://orcid.org/0000-0002-5613-7909}{\includegraphics[width=0.3cm]{orcid_logo.png}} and Wiebke Drenckhan$^{\ast}$\textit{$^{a}$\href{https://orcid.org/0000-0002-3879-4956}{\includegraphics[width=0.3cm]{orcid_logo.png}}}
  \vspace{0.8cm}

\noindent\normalsize{An increasing number of multi-phase systems exploit complex interfaces in which capillary stresses are coupled with solid-like elastic stresses. Despite growing efforts, simple and reliable experimental characterisation of these interfaces remains a challenge, especially of their dilational properties. Pendant drop techniques are convenient, but suffer from complex shape changes and associated fitting procedures with multiple parameters. Here we show that simple analytical relationships can be derived to describe reliably the pressure-deformation relations of nearly spherical elasto-capillary droplets ("droploons") attached to a capillary. We consider a model interface in which stresses arising from a constant interfacial tension are superimposed with mechanical extra-stresses arising from the deformation of a solid-like, incompressible interfacial layer of finite thickness described by a Neo-Hookean material law. We compare some standard models of liquid-like (Gibbs) and solid-like (Hookean and Neo-Hookean elasticity) elastic interfaces which may be used to describe the pressure-deformation relations when the presence of the capillary can be considered negligible. Combining Surface Evolver simulations and direct numerical integration of the drop shape equations, we analyse in depth the influence of the anisotropic deformation imposed by the capillary on the pressure-deformation relation and show that in many experimentally relevant circumstances either the analytical relations of the perfect sphere may be used or a slightly modified relation which takes into account the geometrical change imposed by the capillary. Using the analogy with the stress concentration around a rigid inclusion in an elastic membrane, we provide simple non-dimensional criteria to predict under which conditions the simple analytical expressions can be used to fit pressure-deformation relations to analyse the elastic properties of the interfaces via "Capillary Pressure Elastometry". We show that these criteria depend essentially on the drop geometry and deformation, but not on the interfacial elastiticy. Moreover, this benchmark case shows for the first time that Surface Evolver is a  reliable tool for predictive simulations of elastocapillary interfaces. This opens doors to the treatment of more complex geometries/conditions, where theory is not available for comparison. Our Surface Evolver code is available for download in the Supplementary Materials (see DOI: 10.1039/D1SM01109J). } \\
 \end{@twocolumnfalse} \vspace{0.6cm}

  ]

\renewcommand*\rmdefault{bch}\normalfont\upshape
\rmfamily
\section*{}
\vspace{-1cm}


\footnotetext{\textit{$^{a}$~Institut Charles Sadron, CNRS UPR22 - Universit\'{e} de Strasbourg, Strasbourg, France;  Fax: 33 (0)3 88 41 40 99; Tel: 33 (0)3 88 41 40 43; E-mail: Wiebke.Drenckhan@ics-cnrs.unistra.fr}}
\footnotetext{\textit{$^{b}$~Sorbonne Universit\'{e}s, UPMC Univ Paris 06, CNRS-UMR 7588, Institut des NanoSciences de Paris, 4 place Jussieu, 75005 Paris, France; Email: hohler@insp.upmc.fr} }
\footnotetext{\textit{$^{c}$~Universit\'{e} Gustave Eiffel , 5 Bd Descartes, Champs-sur-Marne, F-77454 Marne-la-Vall\'{e} cedex 2, France. }}
\footnotetext{\textit{$^{d}$~TU Dortmund University, Department of Physics, 44221 Dortmund, Germany }}

\footnotetext{Electronic supplementary information (ESI) available. See DOI: 10.1039/D1SM01109J}







\section{Introduction}
\label{sec:Intro}


The mechanical response of interfaces separating immiscible fluids enters into many fundamental and applied problems of topical interest. Within the current desire to describe complex liquid interfaces \cite{Edwards1991,Rehage_RheoAct_2002,Sagis_RevModPhys_2011,Fuller_SoftMatter_2011,Fuller_2012,Sagis_COCIS_2014,Verwijlen_ACIS_2014}, two scientific communities meet, accustomed  to treating either \textit{drops or bubbles} with \textit{fluid-like} interfaces, or \textit{capsules} and \textit{balloons} whose membranes have \textit{solid-like} mechanical properties. 

\begin{figure}
    \centering
    \includegraphics[width=7cm,keepaspectratio]{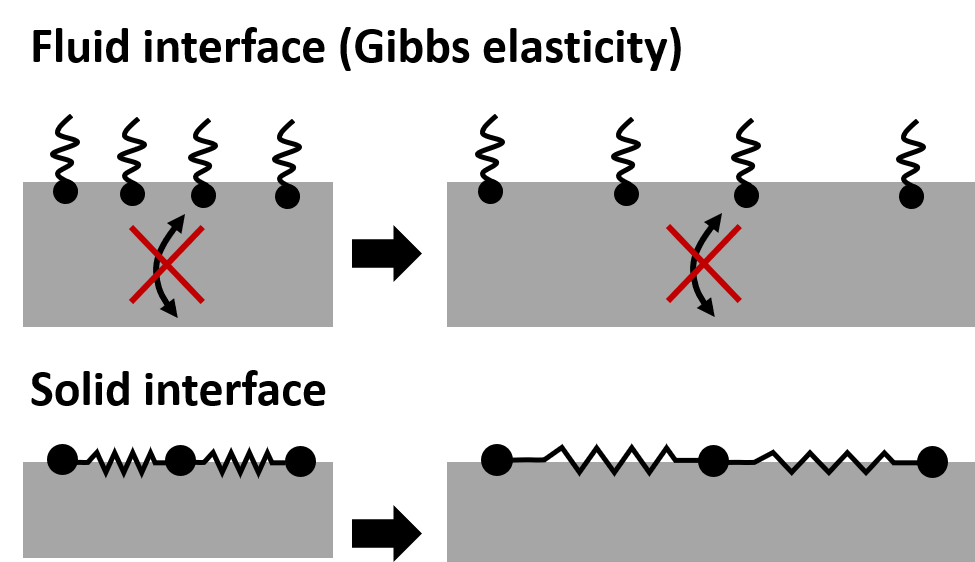}
    \caption{Contrast between the elastic response of fluid and solid interfaces. \textit{Top}: The dilation of the concentration of surfactant molecules adsorbed to fluid interfaces creates an elastic contribution to surface stress. If the exchange of the surfactants with the bulk is inhibited, this response is static and called "Gibbs elasticity". \textit{Bottom}: An elastic stress also appears when a solid like-skin covering the interface is stretched. \label{fig:interfaces}}
\end{figure}

The interfacial tension of complex \textit{fluid interfaces} of drops or bubbles commonly depends on the adsorption of surfactant molecules and on their interactions (top of Fig. \ref{fig:interfaces}). Their interfacial stress is isotropic and in the static limit insensitive to shear deformations. Such fluid systems can present an elastic stress response to dilation in addition to the surface tension. This is commonly called \textit{Gibbs Elasticity} if surfactant exchange between interface and bulk can be neglected. 

The stress in the \textit{solid-like membranes} bounding capsules or balloons (bottom Fig. \ref{fig:interfaces}) strongly depends on both shear deformation and compression away from a stress-free "reference state". These membranes are often thin enough for the elastic bending energy to be negligible compared to those associated with dilation and shear. These "skins" behave like 2D solids, with an elastic response characterised for small deformations by an \textit{interfacial dilational modulus} and a \textit{interfacial shear modulus}. 
\par Like the physics of simple drops/bubbles \cite{Miller1998}, the physics of capsules/balloons \cite{Pozrikidis2003,Mueller_Strehlow_2004,Neubauer_ACIS_2014,Fery_Pol_2007,Sagis2015,Sagis2015a} is now quite well understood. 
However, "intermediate" systems are of increasing interest, which we shall name "droploons" or "bubbloons". Their interfacial properties  combine those encountered respectively in drops/bubbles and capsules/balloons: interfacial tension and solid-like membrane-stresses coexist. Here, the reference state is defined by the absence of a solid-like stress contribution so that only capillary stress is present.  
A multitude of bubbloon- and droploon-like systems have been investigated in the past, involving interfacially active particles, proteins, cross-linked surfactant monolayers, polymer multi-layers, polymer-surfactant mixtures, etc. \cite{Edwards1991,Rehage_RheoAct_2002,Sagis_RevModPhys_2011,Erni2011,Fuller_SoftMatter_2011,Fuller_2012,Sagis_COCIS_2014,Verwijlen_ACIS_2014,Pepicelli_SocRheo_2019}. In most of these systems, liquid- and solid-like elastic contributions are intricately entangled, 
calling for physical models and experimental approaches helping to distinguish and study these contributions. In the following, we provide a very short state of the art of relevant approaches before introducing the one taken for this article. 
Interfacial stresses may in general be of dynamic or static nature and they may present a plastic response depending on deformation history. Here we shall concentrate on the quasi-static response of interfaces. For more details, the reader is referred to recent books and review articles \cite{Edwards1991,Rehage_RheoAct_2002,Miller2009,Sagis_RevModPhys_2011,Fuller_SoftMatter_2011,Fuller_2012,Sagis_COCIS_2014,Verwijlen_ACIS_2014,Pepicelli_SocRheo_2019}. For simplicity, we will also only talk about \textit{droploons} and \textit{liquid/liquid} interfaces, but all derived concepts apply equally to \textit{bubbloons} and \textit{gas/liquid} interfaces.  

The development of dedicated interfacial shear rheometers has enabled reliable measurements of the \textit{interfacial shear modulus} \cite{Miller2009,Kragel2010}. However, the characterisation of the \textit{dilational modulus} remains challenging due to the experimental difficulty of applying an accurately controlled homogeneous dilation to an interface and of assessing the accuracy of the modulus measurement if the deformation is only approximately a homogeneous dilation. \par
Recently, Vermant and coworkers \cite{Verwijlen_ACIS_2014} constructed a special Langmuir trough in which the surface dilation is achieved by the action of twelve fingers arranged circularly. They used this set-up to investigate successfully the static and dynamic dilational response of complex interfaces. However, in order to access the surface stresses, this technique uses a Wilhelmy Balance which introduces potential errors in the measurement due to the influence of the contact line configuration on the Wilhelmy plate. Moreover, the large surfaces required for these measurements are prone to attract impurities, to encourage evaporation and make it challenging to work with liquid/liquid systems. 

\begin{figure}[h!]
\centering
\includegraphics[width=8.5cm]{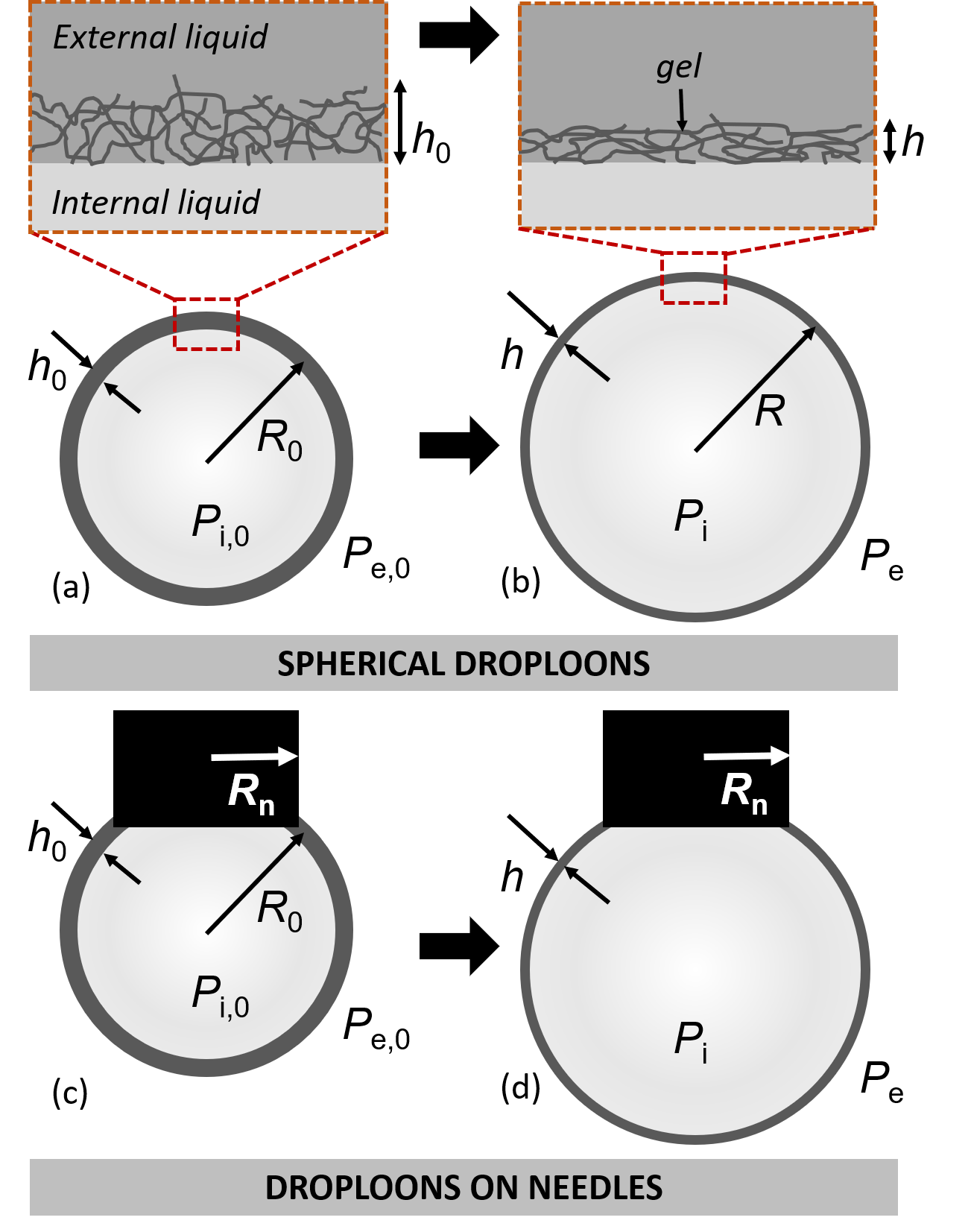}
\caption{Of interest here is the in- and deflation of spherical drops around a reference state of radius $R_0$. These drops are either isolated or attached to a capillary with circular cross-section of radius $R_n$.}
\label{fig:Schemes}
\end{figure}

\par
Since the volume change of a sphere leads to a perfect dilation of its surface, measuring the pressure-radius relation of a small, spherical droploon should be the preferred method to determine the dilational modulus. This has been implemented for capsules using osmotic pressure variations \cite{Gao_EPJ_2001} or acoustic pressure fields \cite{Dollet2019}\footnote{Note that many other techniques have been developed which squeeze initially spherical droploons between two plates, use AFM, spinning drops or investigate the deformation of droploons in controlled flow fields. However, the associated deformations are all a combination of interfacial shear and dilation, making the quantitative analysis extremely complex.}. However, these approaches introduce physico-chemical or technical complexity. It is much more convenient to study the pressure/shape relation of drops held by a capillary with circular cross-section, a technique called "capillary tensiometry" or "pressure tensiometry" when shape or pressure analysis is used, respectively. In the past, it has been used extensively for droploons deformed under gravity, a variant called "capillary elastometry" \cite{Knoche2013,Hegemann2018}. However, the drop shapes are in this case non-spherical with complex interfacial deformations combining shear and dilational components, so that numerical fitting procedures with numerous parameters are required for the data analysis, introducing many uncertainties. Various improvements have been made to these approaches in the past, including improved shape fitting algorithms or combined shape/pressure analysis \cite{Danov_CollIntSci_2015,Nage_l2017}, yet without removing the complexity arising from the non-trivial object shape.

In the aim to identify and validate a quantitative technique for measuring the interfacial dilational modulus, we propose here to use the simplest possible geometry:  an initially spherical droploon attached to a capillary in the absence of gravity. Combining simulations and analytical modelling, we investigate how pressure-deformation relations depend on the interplay between surface tension and solid-like interfacial elasticity. 
Pressure tensiometry of hemi-spherical drops has been exploited in the past \cite{Russev_2008,Kotula_JRheo_2015}, but in all previous work homogeneous isotropic interfacial dilation was assumed. This is an uncontrolled approximation, since such an idealised deformation is incompatible with the boundary imposed by the attachment to the capillary. The exact bubble shape depending on surface tension, the elastic properties of the skin and the gas pressure cannot be calculated analytically. We therefore perform pioneering simulations for this configuration using the Surface Evolver Software - a finite element tool graciously developed and provided by Ken Brakke - in which the combined effects of interfacial tension and specific local mechanical constitutive laws can be implemented. Surface Evolver has already been successfully applied to advance our understanding of systems composed of simple drops \cite{Weaire1999} or of capsules/membranes without surface tension \cite{Bouzidi_CompStruct_2004,Quilliet2016}. Surprisingly, its power has not yet been exploited to perform predictive simulations of droploon-type systems where surface tension and solid-like elasticity are combined. Since direct numerical schemes can be used for the axisymmetric droploon problem (Section \ref{sec:ModelKierfeld}), it provides an ideal benchmark test for the Surface Evolver simulations. The latter will be necessary to predict the response of more complex objects, such as droploon assemblies, where direct numerical schemes will fail.

We treat here a simple model interface, as sketched in the top row of Fig. \ref{fig:Schemes}. We assume it to be composed of a liquid/liquid interface of interfacial tension $\gamma_0$, on which a permanently cross-linked, polymeric gel of thickness $h_0$ is grown. The liquid phase containing the gel is supposed to be a good solvent for the gel, such that the interfacial tension between the gel and the solvent is negligibly small. We furthermore assume that this gel layer is thick enough to be considered a bulk material with bulk shear modulus $G$ and that its mechanical response can be described by a Neo-Hookean model (Section \ref{sec:Theory}). For this purpose, we make the simplifying assumption that the gel can be considered as incompressible, in the sense that its bulk modulus is much larger than its shear modulus. Last but not least, we make the assumption that the gel is dilute enough such that neither its presence nor its deformation modifies the liquid/liquid interfacial tension, thus equal to that of the pure solvent $\gamma_0$.
\par

 After a general introduction to the main theoretical concepts (Section \ref{sec:TheoryFundamentals}), we provide exact analytical relations for the pressure-deformation relations of spherical droploons (Section \ref{sec:TheorySpheres}) and, for the first time, well-matching analytical approximations for droploons on capillaries (Section \ref{sec:TheoryNeedle}). We then show how Surface Evolver can be used to provide reliable simulations of the equilibrium shapes and pressure-deformation relations of this simple physical scenario (Section \ref{sec:SE}), and we show excellent agreement with direct numerical predictions (Section \ref{sec:ModelKierfeld} \cite{Knoche2013,Hegemann2018}). In Section \ref{sec:ResultsDropsNeedles}, we combine theory and simulation to show that the main influence of the capillary results from the change in geometry and not the induced deformation anisotropy. The influence of the capillary on the pressure-volume relationship of a doploon represents a challenging and unsolved theoretical problem because of the interplay of the curved droploon equilibrium shape with the presence of a rigid inclusion, which induces anisotropic elastic deformations of the droploon. We show that this stress anisotropy is strongly localised around the capillary and  provide for the first time analytical relations to estimate the parameter ranges over which the anisotropy at the capillary
  has negligible impact on the pressure-deformation relation, i.e., over which the provided analytical pressure-deformation relations may be used reliably to analyse experiments. We regularly compare with analytical predictions obtained for perfectly fluid interfaces with Gibbs elasticity as a reference. 

 We note that in most experimental systems the interfacial stress may not only depend on deformation but also on the exchange of surfactant molecules between the bulk and the surface or to temperature changes. Sufficiently thick skins may also present a bending stiffness. In addition to an elastic, reversible mechanical response, viscous and plastic behavior is commonly observed.  None of these effects will be considered in the present paper focused on the simplest case of linear and nonlinear 2D elastic skin behavior which is already challenging.



\section{Theory}
\label{sec:Theory}

\subsection{Theoretical framework}

\label{sec:TheoryFundamentals}

Since the recent literature has seen many debates about the physically correct description of the deformation of complex interfaces, we consider it necessary to start here with a fairly general introduction to clarify our point of view before introducing the specific concepts used later in the article.


Interfaces are characterised by the amount of \textit{interfacial free energy per surface area}, that we will denote $f$. If the \textit{interfacial stress} is independent of area changes, the work needed to increase the area by $dA$ is $\gamma dA = f dA $;  $f$ and $\gamma$ are in this case equivalent quantities. However, this is no longer true if the stress and energy density are modified by interfacial area changes. This can be due to interacting surfactant molecules in a fluid-like interface (top of Fig. \ref{fig:interfaces}), or due to a solid, elastic (polymer) skin adsorbed to the interface (bottom of Fig. \ref{fig:interfaces}), or due to a mixture of both.

In this general case, the interfacial stress is no longer necessarily isotropic and its description requires a second rank tensor $\sigma_{ij}$, where $i,j=1,2$ specify components in a 2D cartesian coordinate system locally tangent to the interface.  Assuming that the stresses due to the liquid interfacial tension $\gamma\delta_{ij}$ and those due to the adsorbed elastic skin $\tau_{ij}$ are simply additive one may write \cite{Jaensson_COCIS_2018}
\begin{equation}
    \sigma_{ij} = \gamma\delta_{ij} + \tau_{ij},
    \label{eq:combine}
\end{equation}
where $\delta_{ij}$ is the Kronecker symbol with $\delta_{ij}=1$ if $i=j$ and $\delta_{ij}=0$ otherwise. $\tau_{ij}$ may contain both isotropic and anisotropic contributions, in contrast to $\gamma\delta_{ij}$ which is purely isotropic. The additive decomposition in Eq. (\ref{eq:combine}) should not be taken for granted: if surfactants are cross-linked or co-adsorbed with a polymeric skin, the different contributions to the interfacial stress may be hard to tell apart, not only experimentally but also conceptually. In the present paper, we will not consider this issue further.

  Any measure of  interfacial strain is based on the coordinates of a given interfacial point: $X_i$ in the reference state and $x_i$ after the
deformation ($i=1,2,3$). From these, one may derive the displacement field $U_i(X_i)=x_i - X_i$, where $U_1$ and $U_2$ are the tangential displacements
and $U_3$ the displacement normal to the interface.
For an interface with the two principal radii of curvature in the reference shape $R_{01}$ and $R_{02}$,
displacements give rise to an \textit{infinitesimal} strain tensor \cite{LandauLifshitz}
\begin{equation}
  \epsilon_{ij}=\frac{1}{2}\left(\frac{\partial U_i}{\partial X_j}
    +\frac{\partial U_j}{\partial X_i} +
    \frac{\partial U_3}{\partial X_i} \frac{\partial U_3}{\partial X_j}
  \right)
    +\delta_{ij}\frac{U_3}{2}\left(\frac{1}{R_{01}}+\frac{1}{R_{02}}\right)
    \label{eq:epsilon}
\end{equation}
describing the interfacial 2D strains ($i,j=1,2$). For a spherical surface, the two principal curvature radii are equal ($R_{01}=R_{02}=R_0$) and $\frac{1}{2}(\frac{1}{R_{01}}+\frac{1}{R_{02}})=\frac{1}{R_0}$.
It contains information about the deformation which is
invariant to rotation and translation \cite{LandauLifshitz}.
  Following Kirchhoff's hypothesis \cite{Ventsel2001}, we apply
  classical thin shell approximations, and
  neglect all strains in the plane normal to the interface,
  $\epsilon_{i3}=\epsilon_{3i}=0$ ($i=1,2,3$).
  Both in the Surface Evolver simulations and in the shape equation
  calculus we will employ alternative \textit{finite strain}
  measures, which are introduced below. Their relation to the infinitesimal strain tensor is provided in  Appendix \ref{AppendixA}.


For fluid-like interfaces, stress and strain are isotropic, and in this case scalar quantities of the stress $\sigma$ and the strain $\epsilon$ are useful. They are defined as 
\begin{eqnarray}
    \sigma= \frac{1}{2}(\sigma_{11}+\sigma_{22})  \\  
    \epsilon =\epsilon_{11}+\epsilon_{22}.
    \label{eq:isotropicStressStrain}
\end{eqnarray}
$\epsilon$ is equal to the relative variation of surface area $dA/A$. \par
A rigorous description of \textit{finite strains} can be derived either by considering nonlinear corrections to the kinematics based on the infinitesimal strain tensor \cite{LandauLifshitz,Audoly_2010} or using the displacement gradient tensor \cite{Beatty1987, Mal1991}
\begin{equation}
     F_{ij} = \frac{\partial x_i}{\partial X_j},
     \label{eq:defgrad}
 \end{equation}
  and finally the left Cauchy-Green strain tensor
 \begin{equation}
  B_{ij}= F_{ik} F_{jk},
  \label{eq:defB}
\end{equation}
or the right Cauchy-Green tensor

\begin{equation}
  C_{ij}= F_{ki} F_{kj},
  \label{eq:defC}
\end{equation}
  which extract from $F_{ik}$ information about the strain which is  independent of rotation and translation.  Please note that in this paper we consider right Cauchy Green tensors in 2 and 3 dimensions. To avoid confusions, we denote them respectively as $\mathbf{C}$ and $\mathcal{C}$.
   In this paper, Surface Evolver computes numerically the strain of the surface using the right Cauchy-Green-Tensor, whose explicit expression in the finite element method is derived in the APPENDIX \ref{AppendixA}. For theoretical expressions, however, we will use the left Cauchy-Green tensor, to conform to the commonly used stress-strain expression derived using the Cayley-Hamilton theorem \cite{Macosko}. As stressed by Beatty \cite{Beatty1987}, both tensors have identical principal values (Tr($B_{ij}$)=Tr($C_{ij}$), Tr($B_{ij}^2$)=Tr($C_{ij}^2$), det($B_{ij}$)=det($C_{ij}$)), and are hence equivalent regarding the computation of strain energy.
  
  In Eqs. (\ref{eq:defgrad}) and (\ref{eq:defB}), we use Einstein's summation convention: indices occurring twice should be summed over. \par
 In some models, the Hencky strain is found to be convenient. In the case of an extension that transforms a length $L$ measured in the reference state into a length $L'$, the infinitesimal strain definition in this scalar case would yield $(L-L')/L$ while the Hencky strain is defined as $\ln(L'/L)$. Extensions of the Hencky strain to the tensorial case have been discussed in the literature \cite{Verwijlen_ACIS_2014}.

To build constitutive laws, the strain must be connected to energy density and stress.
Shuttleworth has demonstrated the following general relation between surface stress $\sigma_{ij}$ and surface energy density, assuming constant temperature 
\cite{Shuttleworth_ProcPhysSocA_1950}

\begin{equation}
   \sigma_{ij} = f\delta_{ij} + \frac{\partial f}{\partial \epsilon_{ij}},
    \label{eq:shuttleworth}
\end{equation}
where i,j=1,2. $f$ combines potential energy contributions due to the excess energy of solvent molecules at the interface, adsorbed molecules or elastic potential energy of the skin. 

In the case of fluid interfaces without skins where the stress is isotropic, a scalar model is sufficient. By taking half of the trace of Eq. (\ref{eq:shuttleworth}) and using Eq.s (\ref{eq:isotropicStressStrain}) we obtain the average surface stress, which is equal to the surface tension 
\begin{equation}
     \sigma(\epsilon) =  \gamma(\epsilon) = f +  \frac{\partial f}{\partial {\epsilon}}.
    \label{eq:isotropicshuttleworth}
\end{equation}
 

 
 
For the more general case, we can consider a first order expansion of $\sigma (\epsilon)$ around the reference state yielding
\begin{equation}
    \sigma(\epsilon) =\sigma(0) +K \epsilon,
    \label{eq:K}
\end{equation}

where we have introduced the elastic dilational modulus
\begin{equation}
  K= \left. \frac{\partial f}{\partial\epsilon}\right|_{\epsilon=0}.
    \label{eq:liquid1}
\end{equation}
  

In the spirit of the Hencky strain, the following alternative definition of a dilational modulus, commonly called "Gibbs modulus", is often used \cite{Mysels1961,Kitchener1962}
\begin{equation}
 K_G= \frac{\partial f}{\partial\ln A}.
    \label{eq:Gibbs}
\end{equation}
  For infinitesimal strains, $d \mathrm{ln}A =dA/A = \epsilon $ and both definitions (Eq.s (\ref{eq:liquid1}) and ( \ref{eq:Gibbs})) coincide so that $K = K_G$. For finite strains, there is a distinction between $dA/A$ where  the area $A$ evolves along the deformation and $dA/A_0=\epsilon$ where $A_0$ is the area in the reference state. However, since the Gibbs modulus and the dilational modulus can vary independently as a function of strain, there is no contradiction between the two definitions. Using the Gibbs modulus and assuming its independence of strain amounts to choosing a particular type of constitutive law which appears to describe well some experimental systems  \cite{Salonen_2016,Verwijlen_ACIS_2014}. \\
  
 Let us now turn to interfaces with an adsorbed solid skin. Eq. (\ref{eq:combine}) illustrates our simple hypothesis that the total surface stress is the sum of an interfacial tension and the elastic stress from the skin. To model this latter contribution, we focus on the case where  plastic or viscous response is negligible so that the stress can be derived from a mechanical potential energy. Such materials are called hyperelastic. We focus further on incompressible materials and recall that in this case, the most general constitutive law relating  the three-dimensional elastic stress to deformation can be cast in the form \cite{Beatty1987, Mal1991}
 \begin{equation}
     \sigma^{3D}_{ij} = -p \delta_{ij} + \beta_1 \mathcal{B}_{ij} -  \beta_{-1} \mathcal{B}_{ij}^{-1},
     \label{eq:constitutive}
 \end{equation}
where i,j=1,2,3 and where $p$ is the 3D pressure. The so-called response functions $\beta_1$ and $\beta_{-1}$ depend on the properties of the material and must be expressed as functions of the invariants of the strain tensor to ensure frame invariance. In the simplest case, they are constants leading to what is commonly called the "Mooney-Rivlin" model. It has proven successful in describing many polymers \cite{Macosko,Mueller_Strehlow_2004}. Within this class of models, the case $\beta_{-1} = 0$ is of particular interest. It leads to the so called Neo-Hookean model where $\beta_1$ is equal to the shear modulus $G$ \cite{Beatty1987} so that
\begin{equation}
    \sigma^{3D}_{ij} = -p \delta_{ij} + G\, \mathcal{B}_{ij}.
\end{equation}
This Neo-Hookean model has been derived from a simplified microscopic description of polymer dynamics using statistical mechanics \cite{Larson1998,Mueller_Strehlow_2004}, and it successfully describes the stress response under finite strains. Since for moderate deformations, the Neo-Hooke model remains very close to the Mooney-Rivlin model, it is the method of choice for our simulations.
In the limit of small deformations, the Neo-Hookean model reduces to the well known Hookean model of linear elastic response.
 The 3D mechanical elastic energy density of a Neo-Hookean solid can be expressed as 
 \begin{equation}
 W = \frac{G}{2} (I_\mathcal{B} - 3),
 \end{equation}
 where $I_\mathcal{B}$ is the first invariant of the left Cauchy Green tensor defined in Eq. (\ref{eq:defB}), defined as its trace. This will be useful for the simulations presented in Section 
 \ref{sec:Modelling}.\par


\subsection{Perfectly spherical droploons}
\label{sec:TheorySpheres}

As given in Eq. (\ref{eq:combine}) and sketched in Fig.s \ref{fig:interfaces} and \ref{fig:Schemes}, we assume that the total interfacial stress can be modeled as the sum of  surface tension and and elastic contribution. In the case of fluid-like interfaces, this elastic contribution is given by a Gibbs elasticity. In the case of a solid-like interface, the extra elastic stresses arise from a (Neo-)Hookean skin. 

If the interface is fluid, i.e. only Gibbs elasticity is present, one can integrate Eq. (\ref{eq:Gibbs}) assuming a constant   Gibbs dilational modulus $K_G$. In the limit of negligible gravity (i.e. low density mismatch between the phases or $\Delta\rho gR_0^2/\gamma_0\ll 1$), the reference shape of the drop is spherical and the principal radii of curvature can be assumed to be equal ($R_{01}=R_{02}\equiv R_0$). This gives for a spherical droploon of radius $R$ 
\begin{equation}
    \sigma(A) = \gamma(A) = \gamma_0 + K_{G} \ln{\left( \frac{A}{A_0}\right)} = \gamma_0 + 2K_{G} \ln{\left( \frac{R}{R_0} \right)}.
    \label{eq:GibbsGamma}
\end{equation}

From this, the pressure drop $\Delta P$ across the interface is obtained via the Young-Laplace law
 \begin{equation}
     \Delta P = \frac{2\gamma}{R}.
     \label{eq:Laplace}
 \end{equation}

In the reference state  $R=R_0$ and  $\gamma =\gamma_0$ so that $\Delta P_0 = 2 \gamma_0/R_0$.

To prepare our analysis of solid-like and fluid-like contributions, we introduce the following normalised quantities.
We define an "elastocapillary number"
\begin{equation}
 \alpha = \frac{K}{\gamma_0},
 \label{eq:alpha}
\end{equation}
which compares the dilational elastic modulus $K$ to the interfacial tension $\gamma_0$ of the reference state. $K$ is either due to Gibbs elasticity (denoted $K_G$ in this case) or to a solid-like elasticity, as given later.

For spheres, the stretch $\lambda$ is given by
\begin{equation}
\lambda  = \frac{R}{R_0}.
 \label{eq:areaStretch}
\end{equation}

Moreover, we introduce the normalised interfacial stress

\begin{equation}
\hat{\sigma} =  \frac{\sigma}{\gamma_0}.  \label{eq:NormalisedStress}
\end{equation}
In the case where only Gibbs elasticity is present, the total interfacial stress is therefore given by
\begin{equation}
\hat{\sigma} = \hat{\gamma}=  1 + 2\alpha \ln{\lambda}.
\label{eq:GibbsND}
\end{equation}
 In the small-deformation limit this reduces to 
 \begin{equation}
 \hat{\sigma} = \hat{\gamma} = 1+ 2\alpha (\lambda-1).
 \label{eq:GibbsSmallDef}
 \end{equation}
Whatever the origin of the tension and elastic response  may be, the normalised pressure is obtained using
 
\begin{equation}
    \Delta \hat{P} =  \frac{\Delta P}{\Delta P_0} = \frac{\hat{\sigma}}{\lambda}.
    \label{eq:NormalisedPressure}
\end{equation}
 
Let us now consider solid-like interfaces. For the case of a spherical balloon with initial skin thickness $h_0<< R_0$, starting from Eq. (\ref{eq:constitutive}), Beatty \cite{Beatty1987} derived an analysis valid for any hyperelastic material    
 \begin{equation}
     \Delta P(\lambda)=\frac{2\sigma}{R}=\frac{2 G h_0}{\lambda R_0}\left[ 1-\frac{1}{\lambda^6}\right]\left(\beta_1-\lambda^2\beta_{-1}\right).
 \end{equation}
In the neo-Hookean case this yields the following expression for the stress in the skin 
 \begin{equation}
     \sigma_{Balloon}= G h_0\left[ 1-\lambda^{-6}\right].
 \end{equation}

In several more recent models of non-linear mechanical behavior, nonlinear variations of the response functions with the strain invariants are considered, as reviewed in \cite{Horgan2015,Puglisi2015}. However, for the remainder of this paper we restrict ourselves to the use of the Neo-Hookean model.

We characterised the elastic skin, assumed to be isotropic and incompressible, by its 3D shear modulus $G$. To link it to the 2D dilational modulus, we note that the skin is in a state of plane stress, and that in this case
\begin{equation}
   \epsilon=\epsilon_{11}+\epsilon_{22}=\frac{\sigma_{11}+\sigma_{22}}{2 E} =   \frac{\sigma}{h_0 E} 
\end{equation}
where $E$ is Young's modulus. Here, the biaxial stress in the solid induced by stretching is expressed as a skin tension divided by the skin thickness. In view of Eq. (\ref{eq:K}), this means that $K=E h_0$ in the present case. For incompressible materials $E=3G$, so that for isotropic, small deformations 
\begin{equation}
   K= 3G h_0. 
   \label{eq:SolidModulus}
\end{equation}

In the case of an elastic skin attached to an interface with tension $\gamma_0$ we therefore obtain for the elastocapillary number 
 \begin{equation}
 \alpha = \frac{3Gh_0}{\gamma_0}.
 \label{eq:alphaNH}
 \end{equation}
The total interfacial stress of a spherical neo-Hookean droploon is therefore given by 
 \begin{equation}
 \hat{\sigma} = 1+\frac{G h_0}{\gamma_0}(1-\lambda^{-6})=1+\frac{\alpha}{3}(1-\lambda^{-6}).
 \label{eq:NeoHookeTension}
 \end{equation}
In the small deformation limit one obtains the prediction of the linear elastic Hooke model
 \begin{equation}
 \hat{\sigma} = 1+ 2\alpha(\lambda-1),
 \label{eq:HookeTension}
 \end{equation}
 which is identical to Eq. (\ref{eq:GibbsSmallDef}). This result shows that in the limit of isotropic and small deformations both Gibbs elasticity and Neo-Hookean elasticity lead to a linear elastic response captured by Hooke's law in two dimensions with a compression modulus $K_G = 3Gh_0$.\par
 
 \begin{table*}
\renewcommand{\arraystretch}{2}
    \centering
    \begin{tabular}{|c|c|c|c|}
    \hline
Sphere model  & Normalised surface stress $\hat{\sigma}$ & Critical stretch $\lambda_{A,c}$ & Stretch at maximum pressure $\lambda_{A,m}$ \\ \hline
Gibbs (liquid) &  $1 + \alpha \ln{\lambda_A}$ & $ \exp{\left(-\frac{1}{\alpha}\right)}$ & $ \exp{\left(2-\frac{1}{\alpha}\right)}=e^2\lambda_{A,c}$\\ \hline
Neo-Hooke (solid) &  $1 + \frac{\alpha}{3} (1-\lambda_A^{-3})$ &  $ \left( \frac{\alpha}{\alpha+3} \right) ^{1/3} $ & $\left( \frac{7\alpha}{\alpha+3} \right)^{\frac{1}{3}} =7^{\frac{1}{3}}\lambda_{A,c}$ \\ \hline
Hooke &  $1 + \alpha(\lambda_A-1) $ & $ \left( 1-\frac{1}{2\alpha} \right)^2$ (for $\alpha>0.5$) & no maximum \\
\hline
    \end{tabular}
    \caption{Summary of the normalised expressions for the normalised surface stress $\hat{\sigma}=\sigma/\gamma_0$; the critical stretch $\lambda_{A,c}$ at which the pressure changes sign; and the stretch at maximum pressure $\lambda_{A,m}$ for the Gibbs, Neo-Hooke and Hooke model.}
    \label{tab:models}
\end{table*}
 
Eq. (\ref{eq:HookeTension}) shows that for $\alpha > 1/2$, an extensional stretch induces a positive total surface stress, acting as a restoring force while for $\alpha < 1/2$ an extensional stretch yields a negative total stress which favors further deformation. Analogous tendencies are predicted for compression. The condition $\alpha = 1/2$ has therefore received particular attention and is often called the "Gibbs criterion" since the physical response of a system may change fundamentally around this value. This is known, for example, for the case of bubble dissolution and foam coarsening \cite{Stocco2009,Salonen_2016}.

In the case of spheres, it is natural to express interfacial stresses and curvatures via the radial stretch $\lambda$. However, for more general surfaces, the relationship between both depends on the geometry of the surface. In this case it is more appropriate to express the dilational stresses via the area stretch $\lambda_A = A/A_0$. 
For spheres, the relationship between area and radial stretch is simply 
\begin{equation}
\lambda  = \frac{R}{R_0} = \left( \frac{A}{A_0} \right)^{1/2} = \lambda_A^{1/2}.
 \label{eq:radiusStretch}
\end{equation}
In Table \ref{tab:models} we summarise the interfacial stresses for the Gibbs, Neo-Hookean and Hookean model expressed via their area stretches, together with some critical stretches which are discussed in Section \ref{sec:ResultsSpheres}. In the following we will use those relations.

\subsection{Droploons on capillaries}
\label{sec:TheoryNeedle}
Let us now consider droploons attached to capillaries with circular cross-section of radius $R_n$ (Fig. \ref{fig:Schemes}). In this case one geometrically removes a cap of radius $R_n$ from the droploon and fixes the perimeter of the resulting circular hole to the end of the capillary. For fluid interfaces with Gibbs elasticity, the interfacial stresses are isotropic and constant everywhere in the interface, even if the droploon is inflated or deflated. Hence, the droploon shapes remain spherical sectors and, as we show below, all pressure-deformation relations can be calculated analytically, giving useful insight into the impact of the geometry change. In the case of interfaces with a solid skin, this is much less straightforward. Fixing the interface points on the capillary boundary induces shear deformation in the vicinity of the capillary upon inflation or deflation and hence deviations from the shape of a perfect sphere. The presence of the capillary in the case of a solid-like skin therefore combines a geometrical impact (as for the Gibbs elasticity) with one of a non-isotropic deformation. Both contributions are coupled and their relative importance depends on the capillary number $\alpha$, on the deformation $A/A_0$ and on the capillary-to-drop size ratio $R_n/R_0$.\\
Let us assume in the following that shear stresses remain negligible and that we can estimate the droploon shape by spherical sectors derived from perfect spheres of radius $R$ from which a cap of radius $R_n$ is removed, as depicted in Fig. \ref{fig:Schemes}. The interfacial  area $A$ is then given by 
\begin{equation}
    \begin{split}
        A(R) & = 2\pi R^2\left(1 \mp \sqrt{1-\left(\frac{R_n}{R}\right)^2}\right),
    \end{split}
    \label{eq:interfacialarea01}
\end{equation}
where the two signs correspond to droploons larger than a hemisphere  ("+") or smaller than a hemisphere ("-").The latter geometry introduces a major difference between drops with and without capillaries: the radius of the drop \textit{increases} upon further deflation from the hemisphere. This changes dramatically the pressure-deformation relation, which is why we will exclude this case in the remaining discussion.\\
Eq. (\ref{eq:interfacialarea01}) can be used to relate the area stretch $\lambda_A$ and the radial stretch $\lambda$ via

\begin{equation}
    \begin{split}
        \lambda  = & \;\lambda_A^{1/2}\frac{1+\sqrt{1-\left(\frac{R_n}{R_0}\right)^2}}{\sqrt{2\left[1+\sqrt{1-\left(\frac{R_n}{R_0}\right)^2}\right] -\left(\frac{R_n}{R_0}\right)^2\frac{1}{\lambda_A}}} \\
         = &  \;\lambda_A^{1/2}  \; \mathpzc{f}\left( \frac{R_n}{R_0},\lambda_A\right).
        \end{split}
    \label{eq:GeometricalCorrection}
\end{equation}

i.e. when comparing with the full sphere expression of Eq. (\ref{eq:areaStretch}), the presence of the capillary introduces a correction factor $\mathpzc{f}\left( \frac{R_n}{R_0},\lambda_A\right)$ to the relationship between the radial and the area stretch. 

For a given area stretch $\lambda_A$ - which is experimentally and computationally more easily accessible than the radial stretch $\lambda$ - we can then rewrite the pressure-deformation relation as

\begin{equation}
   \Delta \hat{P}=\frac{\hat{\sigma}(\lambda_A)}{\lambda}=\frac{\hat{\sigma}(\lambda_A)}{\lambda_A^{1/2}} \mathpzc{f}^{-1} = \Delta \hat{P}_S \mathpzc{f}^{-1},
    \label{eq:PressureDefNeedle}
\end{equation}
where $\Delta \hat{P}_S$ is the pressure of the sphere with the same area stretch and the interfacial stress $\hat{\sigma}$ is given in Table \ref{tab:models} for the different models. Hence, in the approximation of negligible shear contributions, the capillary may be considered to impose a simple geometrical correction on the pressure-deformation relation which depends only on the capillary size $\frac{R_n}{R_0}$ and the area stretch $\lambda_A$.  In the case of fluid-like interfaces (Gibbs elasticity), Eq. (\ref{eq:PressureDefNeedle}) is accurate, while in the case of solid-like interfaces (Neo-Hooke \& Hooke), this is an approximation. We shall see in Section \ref{sec:ResultsDropsNeedles} that this remains nevertheless an excellent approximation over a wide range of parameters.

Here we have chosen to express the pressure-deformation relations in terms of area stretch $\lambda_A$ since it simplifies comparison with simulations and experiments. One may also choose to express them in terms of radial stretch $\lambda$. In this case it is the expression of the interfacial stress $\hat{\sigma}$ which needs to be modified, leading to more complex expressions. We provide these relations for the interested reader in Annex \ref{annex:PressDefNeedle}.


\section{Numerical modelling}
\label{sec:Modelling}
\subsection{Surface Evolver simulations}
\label{sec:SE}
\label{Subsec:SEPrinciple}

Surface Evolver\cite{Brakke1992} is a widely used software  that determines the equilibrium structure of systems containing several fluid phases separated by interfaces. It uses the principle that in equilibrium, the interfacial energy must be minimal under the constraints imposed by boundary conditions. Examples of this are foams where the volume of each bubble is fixed \cite{Buffel2014,Weaire2017,Hohler2017,Ginot2019}.  Surface Evolver  can also be used to model elastic membranes \cite{Bouzidi_CompStruct_2004,Quilliet2016}.


In Surface Evolver simulations, interfaces are represented as meshes of triangular facets whose energy is evaluated. Most previous studies on bubble or drop shapes  focus on systems where this energy is proportional to the interfacial area, the proportionality factor being the surface tension $\gamma$. Additionally to this contribution, Surface Evolver simulations can also take into account an elastic energy induced by the deformation of each facet, simulating an elastic skin.  Several constitutive laws are implemented in the Evolver Software and can be used: Hooke's law describing linear elastic response, as well as the non-linear Saint-Venant or Neo-Hooke's law\cite{Bouzidi_CompStruct_2004}. In the work reported here, we use Neo-Hooke's law introduced in Section \ref{sec:TheoryFundamentals}. We implement, for the first time to our knowledge, an interface with both surface tension and neo-Hooke interfacial elasticity. As a first implementation, we thoroughly compare Surface Evolver results to the numerical solution of the shape equations (Section \ref{sec:ModelKierfeld}), and ensure that it provides physically sound results in the investigated range of parameters.

In contrast to fluid interfaces where the interfacial area uniquely determines the energy, the energy of elastic skins depends on their deformation with respect to a reference state. The reference state of an interface element is given by a shape with zero interfacial elastic stress. This state is encoded in the reference positions of the facet vertices. The implementation of elastic stress in the framework of the Surface Evolver requires an expression of the facet deformation energy for arbitrary large strains, given as a function of the vertex positions. A detailed presentation of this feature and the implementation of elastic energy in the Surface Evolver has not been published so far to our knowledge. We therefore provide this information in the Appendix \ref{AppendixA} to clarify for the interested reader how exactly the software operates. Here we shall concentrate on a very general description of the approach.

Our Surface Evolver calculations simulate an experiment where a bubble or drop is inflated at the tip of a cylindrical hollow capillary inserted into a liquid, as illustrated in Fig. \ref{fig:DroploonSimulations}. In the first step, we need to obtain a physically correct reference shape for a drop without interfacial elasticity.  For this purpose, an initially very coarse mesh is attached to a cylindrical boundary representing the capillary. The interfacial area is then minimised for the given drop target volume assuming that interfacial energy is due only to a uniform and constant surface tension\footnote{This could represent a physical system where the elastic skin forms progressively at an initially "naked" interface}. Successive refinements and energy minimisations of the mesh are then performed to simulate the drop shape and the pressure in the reference bubble accurately. When the relative variation of total interfacial energy $|E^{n+1}-E^{n}|/E^n$ remains smaller than $10^{-8}$ over 100 iteration steps we consider that convergence has been achieved.\\
In the second step of the simulation, an elastic skin is added to the drop surface of the obtained reference state, so that initially there is no elastic stress. Numerically, it consists in saving the current positions $\{\vec{X}_i\}$ of the vertices as their reference positions, and setting a non-zero elastic modulus value for the interfacial energy computation for further minimisation iterations. How reference and current positions are used for deformation computation is detailed in Appendix \ref{AppendixA}.\\
The third step consists in inflating or deflating this droploon up to a new volume where mechanical equilibrium is again established via progressive mesh  relaxation. Frequent merging of facets significantly smaller than average and refinement of facets larger than average hastens convergence whilst avoiding to trap the system in local energy minima. These operations are all performed by Surface Evolver in-built routines as part of a standard energy minimisation procedure. When the mesh management and energy minimisation have converged ($|E^{n+1}-E^{n}|/E^n<10^{-8}$), the elastic stress in the skin, the pressure in the bubble and the bubble shapes are recorded.\\


\subsection{Numerical integration of the shape equations} 
\label{sec:ModelKierfeld}

We solve for the shape and stress/strain profile of an axi-symmetric capsule by 
numerically integrating the \emph{shape equations} \cite{Hegemann2018,Knoche2013}. Because we impose axial symmetry, 
the droploon can be parametrised as a single arc with
arc length $s$ and arc angle $\Psi$. The transformation from arc length parametrisation to cylindrical coordinates $\{r, \phi, z\}$ gives the first two shape equations
\begin{equation}
    \label{eqn:shape_eqn_rz} \frac{\mathrm{d}r}{\mathrm{d}s} = \cos\Psi ~~~\text{and}~~~
     \frac{\mathrm{d}z}{\mathrm{d}s} = \sin\Psi \,.
\end{equation}
The remaining shape equations, needed to close the set of partial differential equations, take into account the constitutive material law and reflect the force balance at every point along the arc $s$. They are derived by searching for the stationary solutions of the appropriate energy functional.

\begin{figure}[h]
    \centering
    \includegraphics[width=7cm,keepaspectratio]{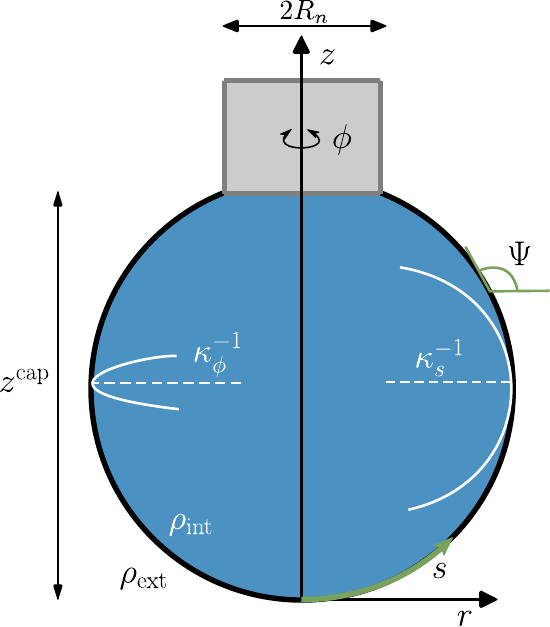}
    \caption{A pendant droploon parametrised in arc-length $s$ and arc-angle
      $\Psi$.
      \label{fig:parametrization}}
\end{figure}

In the experimentally relevant setting we control either the droploon volume or the mechanical pressure at the capillary inlet. Thus, the appropriate energy functional is the enthalpy
\begin{equation}
    H = \int \mathrm{d}A_0 \, W_{2D} + \int \mathrm{d}A \, \gamma_0 - \int \mathrm{d}V \, \Delta P \,,
    \label{eqn:enthalpy_functional}
\end{equation}
with a contribution from the surface energy $W_{2D}$, measured 
with respect to the undeformed area $A_0$, from the surface tension $\gamma_0$ and from the volumetric work 
against a pressure difference $\Delta P$.
We find the stationary states of the enthalpy $H$ of Eq. (\eqref{eqn:enthalpy_functional}) via the first 
variation, $\delta H = 0$ (see \cite{Knoche2013, Hegemann2018} for details), leading to the shape equations
\begin{align}
    \label{eqn:shape_eqn_psi} \frac{\mathrm{d} \Psi}{\mathrm{d}s} &= \kappa_s = \frac{1}{\sigma_s} \left(\Delta P - 
    \kappa_\phi \sigma_\phi \right)
     \,, \\
    \label{eqn:shape_eqn_taus} \frac{\mathrm{d} \sigma_s}{\mathrm{d}s} &= \frac{\cos\Psi}{r} \left( \sigma_\phi - \sigma_s \right) \,,
\end{align}
where ($\kappa_s,\kappa_{\phi}$) and ($\sigma_{s},\sigma_{\phi}$) are the meridional and  circumferential  curvatures and surface stresses, respectively. The curvatures are given by 
$\kappa_\phi = {\sin \Psi}/{r}$ and 
$\kappa_s = {\mathrm{d}\Psi}/{\mathrm{d}s}$.
Note that the shape equations \eqref{eqn:shape_eqn_rz},  \eqref{eqn:shape_eqn_psi} and \eqref{eqn:shape_eqn_taus} still require a constitutive material law for closure.
At this point, no detailed knowledge about the 2D surface energy functional $W_{2D}$ is required, as we
define 
\begin{equation}
    \sigma_{s, \phi} = \frac{1}{\lambda_{\phi, s}}\left(\frac{\partial W_{2D}}{\partial \lambda_{s, \phi}} 
    + \frac{\partial (\gamma_0 \lambda_s \lambda_\phi)}{\partial \lambda_{s, \phi}}\right) \,,
    \label{eqn:stress_energy_functional}
\end{equation}
where $\lambda_s$ and $\lambda_\phi$ are the meridional and circumferential
stretch ratios of the droploon.  The shape equations \eqref{eqn:shape_eqn_rz},
\eqref{eqn:shape_eqn_psi} and \eqref{eqn:shape_eqn_taus}
are written in terms of the arc length $s$ of the deformed 
shape. For the numerical solution we reparametrise in terms of the
\emph{undeformed} arc length coordinate $s_0$ of the original undeformed
shape by using the relation
$\mathrm{d}s / \mathrm{d}s_0 = \lambda_s$, which is necessary in order to
gain access to the meridional stretches $\lambda_s$.
The circumferential stress $\lambda_\phi = r/r_0$
is given by the ratio of undeformed and deformed 
radial coordinate.

The surface energy $W_{2D}$ accounts for the material specific model and can incorporate various effects, such as film thinning. To express the constitutive equation in terms of our parametrisation we write the right 2D Cauchy-Green tensor, discussed in Section \ref{sec:Modelling} and in Appendix \ref{AppendixA}, as  
\begin{equation}
    \mathbf{C} = \mathrm{diag}(\lambda_s^2, \lambda_\phi^2)\,.
\end{equation}
For a two-dimensional Neo-Hookean elastic material the surface energy is given by Eq. \eqref{eq:2D energy density final} from the Appendix \ref{sec:energy}
\begin{equation}
    W_{2D} =\frac{G h_0}{2} \left( \mathrm{Tr}\mathbf{C} + \mathcal{C}_{33} + \frac{G}{\Lambda} \mathcal{C}^2_{33}\right).
\end{equation}
with 3D Lamé parameters $G$ and $\Lambda$.
Here, $\mathbf{C}$ is the 2D Cauchy-Green tensor describing deformations within the surface, while 
$\mathcal{C}_{33}$ is the component of the 3D  Cauchy-Green tensor describing normal (thickness) deformations of the elastic skin. Requiring the absence of normal stresses, $\mathcal{C}_{33}$ becomes a function of $G/\Lambda$ and $\mathrm{det}\mathbf{C}$ as derived in Appendix \ref{sec:energy}.

From this surface energy,  we extract the constitutive law needed to close the shape equations using Eq.\ \eqref{eqn:stress_energy_functional},
\begin{equation}
    \sigma_{s, \phi} = G h_0 \left( \frac{\lambda_{s, \phi}}{\lambda_{\phi, s}} - \frac{\mathcal{C}_{33}}{\lambda_s \lambda_\phi} \right) + \gamma_0 \,.
\end{equation}
In the following, we focus on the incompressible limit $G / \Lambda \ll 1$, where $\mathcal{C}_{33} \approx 1 / \mathrm{det}\mathbf{C} = 1 / \lambda_s^2 \lambda_\phi^2$. 

For a given undeformed shape (described by a function $r_0(s_0)$),
the shape equations, along with the constitutive equations, are numerically integrated from the apex ($s=0$) to the attachment point at the capillary ($s=L$) using a Runge-Kutta scheme, paired with a shooting algorithm to satisfy the boundary conditions
\begin{equation}
 r(s = 0) = z(s = 0) = \Psi(s = 0) = 0~~~\text{and}~~~
 r(s = L) = R_n  .
\end{equation}
In the shooting procedre, 
we prescribe an apex stress $\sigma_s(s = 0)$ and iteratively search for
a pressure drop $\Delta P$ satisfying the attachment boundary condition at the
capillary. Moreover, we restrict the prescribed apex stresses to the
physically relevant ones for our context giving $\sigma_s(s = 0) > 0$ (no
compressive stresses), and do not exceed the maximal possible apex stress
allowed by the constitutive equations,
$\sigma_{s, \phi}(s = 0)^{\text{max}} = G h_0 + \gamma_0$.


\section{Results}
\label{sec:Results}


In Section \ref{sec:ResultsSpheres} we compare the theoretical predictions of the different elastic laws in Eqs. \eqref{eq:GibbsND}, \eqref{eq:alphaNH} and \eqref{eq:NeoHookeTension}, and the results obtained from Surface Evolver simulations. In Section \ref{sec:ResultsDropsNeedles}, we compare the numerical simulations to the analytical predictions where the needle is treated as a geometrical perturbation truncating an isotropic droploon (Section \ref{sec:TheoryNeedle}). These two results are compared to the direct numerical predictions (Section \ref{sec:ModelKierfeld}), which account both for the geometrical perturbation and the anisotropy imposed by the needle. Finally, we quantify the perturbation of the pressure induced by the needle, and show that it can be in large part explained by the geometrical perturbation. In the last step, we use the direct numerical predictions to quantify the importance of anisotropic stretches, and provide experimentalists with guidelines to predict the parameter ranges over which the influence of the capillary (shape change and/or stress anisotropy) can be neglected.

\subsection{Spherical droploons}
\label{sec:ResultsSpheres}

\begin{figure*}[h]
\centering
\includegraphics[width=13cm,keepaspectratio]{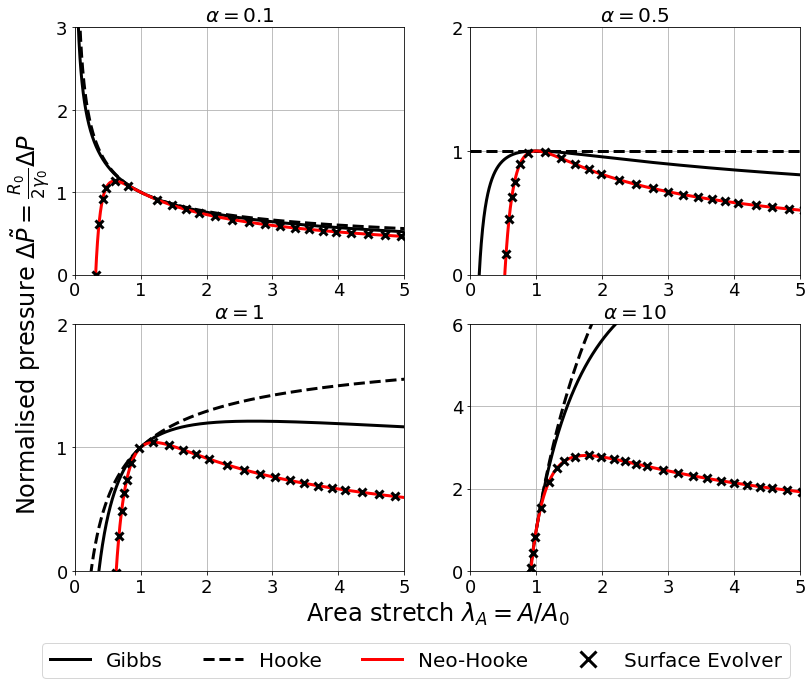}
\caption{Normalised pressure as a function of area stretch $\lambda_A$ for spherical droploons whose skin elasticity is described by Gibbs', Neo-Hooke's or Hooke's law. Four characteristic elastocapillary number values ($\alpha = 0.1$, $0.5$,$1$,$10$) are investigated. The data obtained by Surface Evolver simulations are obtained assuming Neo-Hookean elasticity.  }
\label{fig:PressureDeformationSPhere}
\end{figure*}
 
\begin{figure}[h!]
\centering
\includegraphics[width=9.5cm]{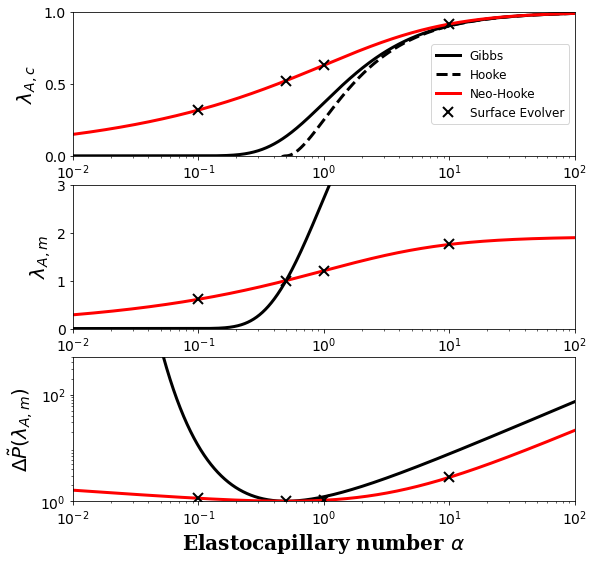}
\caption{Variation of characteristic features (critical area stretch $\lambda_{A,c}$, stretch $\lambda_{A,m}$ at maximum pressure and maximum pressure $\Delta \hat{P}(\lambda_{A,m})$) with the elastocapillary number $\alpha$ predicted for droploons with skins presenting Gibbs, Neo-Hookean and Hookean elasticity. Surface Evolver simulations are performed for the Neo-Hooke case. }
\label{fig:SummFeaturesSphere}
\end{figure}

We run Neo-Hookean Surface Evolver simulations (Section \ref{sec:SE}) for spheres with four different elastocapillary numbers ($\alpha =$ 0.1, 0.5, 1, 10) imposing inflation and deflation while recording the normalised pressure difference $\Delta \hat{P}$. The results are shown in Fig. \ref{fig:PressureDeformationSPhere} as a function of area stretch $\lambda_A$ along with the theoretical predictions for the Gibbs, Hooke and Neo-Hooke models provided in Section \ref{sec:TheorySpheres}. 

The simulations show excellent agreement with the Neo-Hookean theory over the full range of investigated deformations. As expected and discussed in Section \ref{sec:TheorySpheres}, all three models coincide in the small deformation limit  $\lambda_A \approx 1$. However, for deformations of a few percent, the three models already show very pronounced differences, indicating the importance of choosing the physically most realistic model for the interpretation of pressure-deformation relations. 

For non-zero $\alpha$, in the case of the Gibbs and Neo-Hookean elasticity, the initially monotonously decreasing Young-Laplace-like behaviour is replaced by a pressure-deformation relation with a well-pronounced pressure maximum $\Delta\tilde{P}(\lambda_{A,m})$ at a characteristic stretch $\lambda_{A,m}$. Upon deflation ($\lambda_A<1$), this leads to the  apparition of a critical stretch $\lambda_{A,c}$ at which the pressure difference is zero, and beyond which it becomes negative. This point corresponds to elastic instabilities of compressed interfaces, which manifest themselves in buckling phenomena \cite{LandauLifshitz,Zoldesi1998,Sacanna2011}. A proper handling of this range requires to take into account the bending energies of the interfaces. Since this is neither of interest here, nor implemented in our simulations, we stay away from the buckling range in our analysis.

The variation of $\lambda_{A,c}$, $\lambda_{A,m}$ and of the pressure difference $\Delta\tilde{P}$ at $\lambda_{A,m}$ with elastocapillary number $\alpha$ for the different models are shown in Fig. \ref{fig:SummFeaturesSphere}. The corresponding analytical expressions are given in Table \ref{tab:models}. They put in evidence clear differences between Gibbs, Hookean and Neo-Hookean models. In comparison to Gibbs elasticity, the Neo-Hookean critical and maximal stretches vary only mildly with $\alpha$. The Surface Evolver results again agree very well with theory. The critical stretch for Hooke's model appears when the elastocapillary number crosses the Gibbs criterion $\alpha=0.5$. The Gibbs critical stretch tends exponentially towards 0, as $\lambda_{A,c}=\mathrm{exp}(-1/\alpha)$. In the limit of large $\alpha$, the critical stretches all converge towards $\lambda_{A,c}=1$, that is, a shell so rigid that it buckles as soon as compressed. 
Hooke elasticity does not predict a local pressure maximum at any elastocapillary number. But it predicts an interesting deformation-independent pressure for $\alpha=0.5$, i.e. at the "Gibbs criterion". Gibbs and Neo-Hooke, on the other hand, have a maximal pressure stretch increasing with $\alpha$. In particular, at the Gibbs criterion $\alpha=0.5$, the maximal pressure is reached at null deformation ($\lambda=1)$. Lower elastocapillary numbers move $\lambda_{A,m}$ to the compression regime ($\lambda_{A,m}<1$), while $\alpha>0.5$ shift $\lambda_{A,m}$ to the dilation regime ($\lambda_{A,m}>1$).
The most remarkable features of the elastocapillary transition (onset of significant critical stretch, variation of the maximal pressure stretch) occur for elastocapillary numbers between $0.1$ and $10$. For this reason, we expose in this article results for $\alpha=0.1$, $1$ and $10$, so as to span two decades of elastocapillary numbers. Because of its history as the Gibbs criterion and its pivot point between capillarity and elasticity, $\alpha=0.5$ will also be represented.

\subsection{Droploons on capillaries}
\label{sec:ResultsDropsNeedles}

\begin{figure}[h!]
\centering
\includegraphics[width=8.5cm,keepaspectratio]{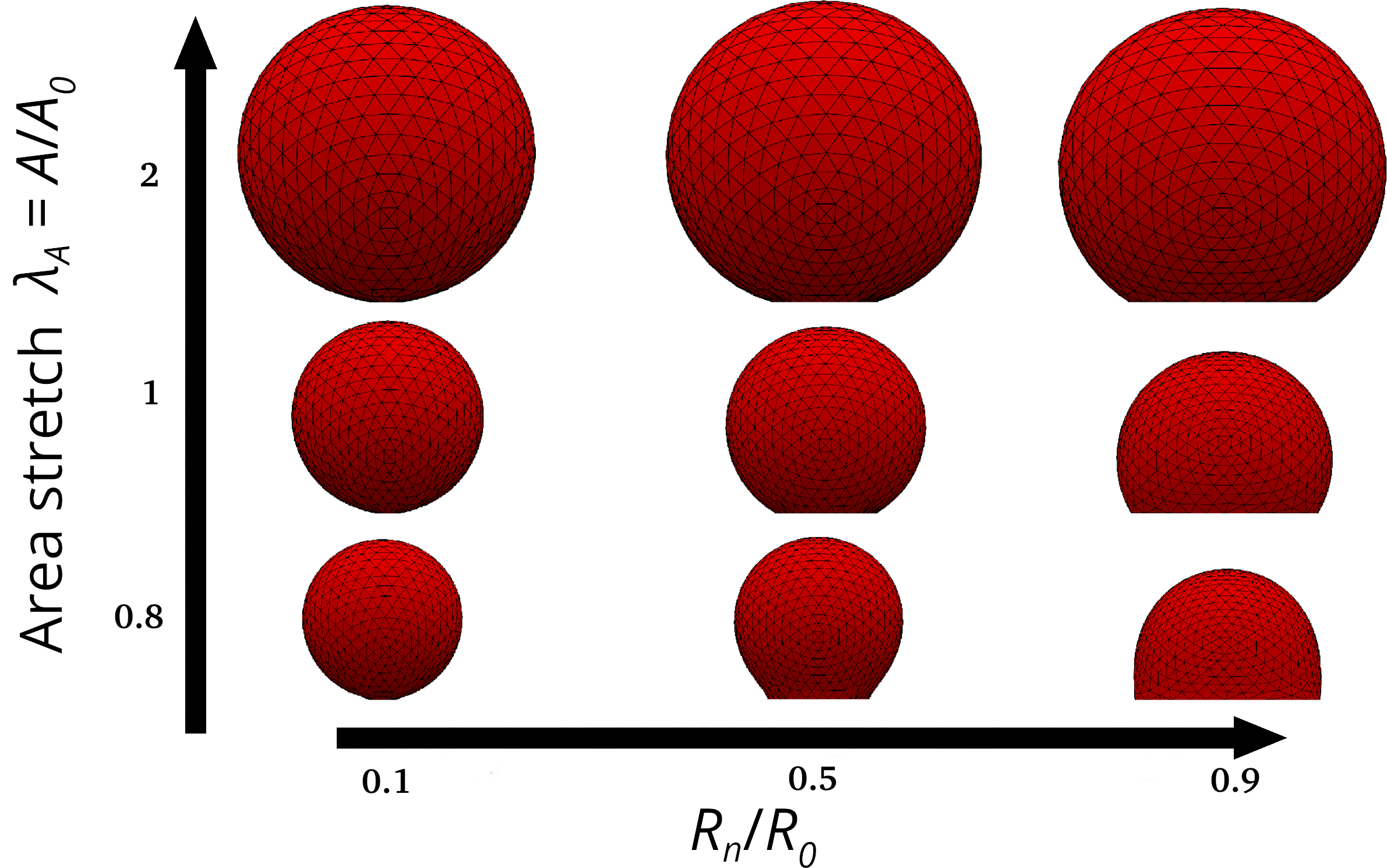}
\caption{Examples of neo-Hookean droploons at different area stretches and capillary ratios $R_n/R_0$ obtained for $\alpha = 0.5$ using Surface Evolver.}
\label{fig:DroploonSimulations}
\end{figure}

In a second step, we run Surface Evolver simulations of pendant droploons attached to a capillary with circular cross-section of radius $R_n$ (Fig. \ref{fig:Schemes}). The droploons are inflated and deflated while their interfacial area and inner pressure are recorded (Section \ref{sec:SE}). Three  ratios between the capillary radius $R_n$  and the radius $R_0$ of the droploon in the reference configuration are used: $R_n/R_0$ = 0.1, 0.5 and 0.9. Representative examples of obtained droploon shapes are shown in Fig. \ref{fig:DroploonSimulations} for three characteristic area stretches ($\lambda_A=$ 0.8, 1, 2) for the case of $\alpha = 0.5$. 

\begin{figure*}
\centering
\includegraphics[width=15cm,keepaspectratio]{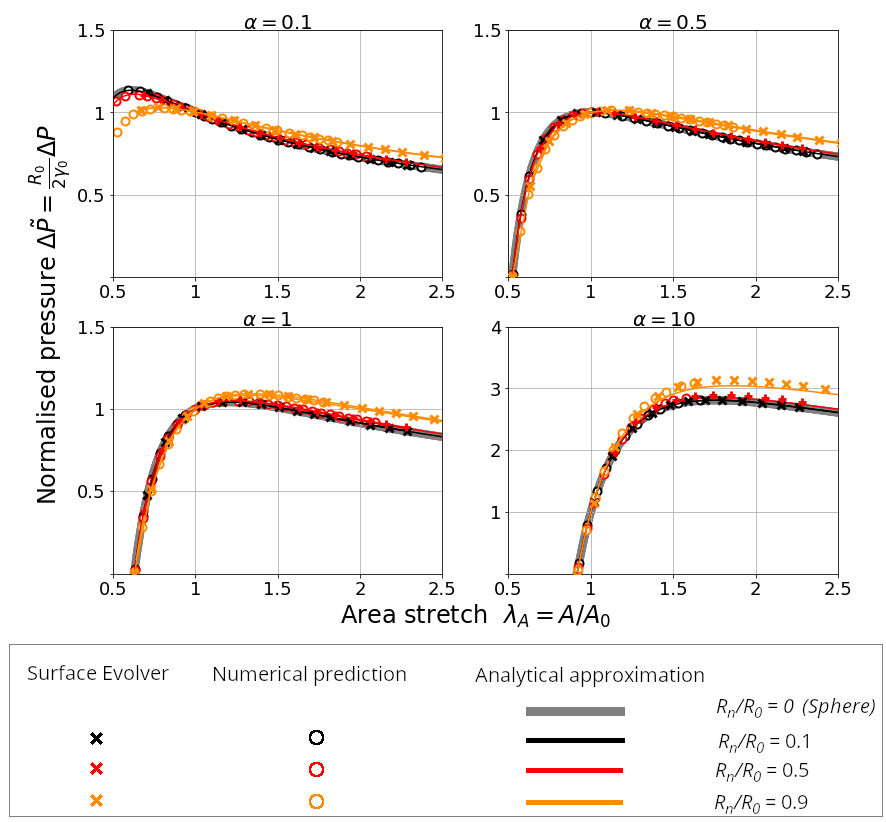}
\caption{Normalised pressure as a function of area stretch $\lambda_A$ of Neo-Hookean droploons on capillaries for three ratios of capillary and initial droploon radius ($R_n/R_0 = 0.1, 0.5, 0.9$), and four characteristic elastocapillary numbers ($\alpha = 0.1, 0.5,1,10$). Surface Evolver simulations are compared with direct numerical predictions (Section \ref{sec:Modelling}) and with the analytical expression of Eq.\ (\ref{eq:NeoHookeTension}) using a simple geometrical correction to the perfect sphere theory.}
\label{fig:PressureDeformationSPhereNeedle}
\end{figure*}

In Fig. \ref{fig:PressureDeformationSPhereNeedle} we show the obtained pressure-deformation relations for the elastocapillary numbers $\alpha = 0.1, 0.5, 1, 10$. Along with the Surface Evolver results (crosses) we plot results obtained by direct numerical predictions (empty circles) using the Neo-Hookean shape equations for the same set of parameters (Section \ref{sec:ModelKierfeld}). The excellent agreement between both for all elastocapillary numbers, capillary radii and deformations demonstrates the reliability of Surface Evolver simulations for such systems. 

The solid lines shown in Fig.  \ref{fig:PressureDeformationSPhereNeedle} correspond to the analytical approximation given in  Eq. (\ref{eq:PressureDefNeedle}) which models droplets as spherical sectors covered with a Neo-Hookean skin. The agreement is excellent in the whole deformation range for all capillary sizes and elastocapillary numbers. This means that in this parameter range the deviation from the predictions for spherical droploons without any capillary (gray line in Fig. \ref{fig:PressureDeformationSPhereNeedle}) are essentially a result of the associated change of the geometry induced by the capillary, rather than due to the shear deformation in the vicinity of the capillary. Deviations from the simple model set in only for large capillary sizes ($R_n/R_0=0.9$) and large elastocapillary numbers ($\alpha = 10$). 

\begin{figure*}
    \centering
    \includegraphics[width=15cm]{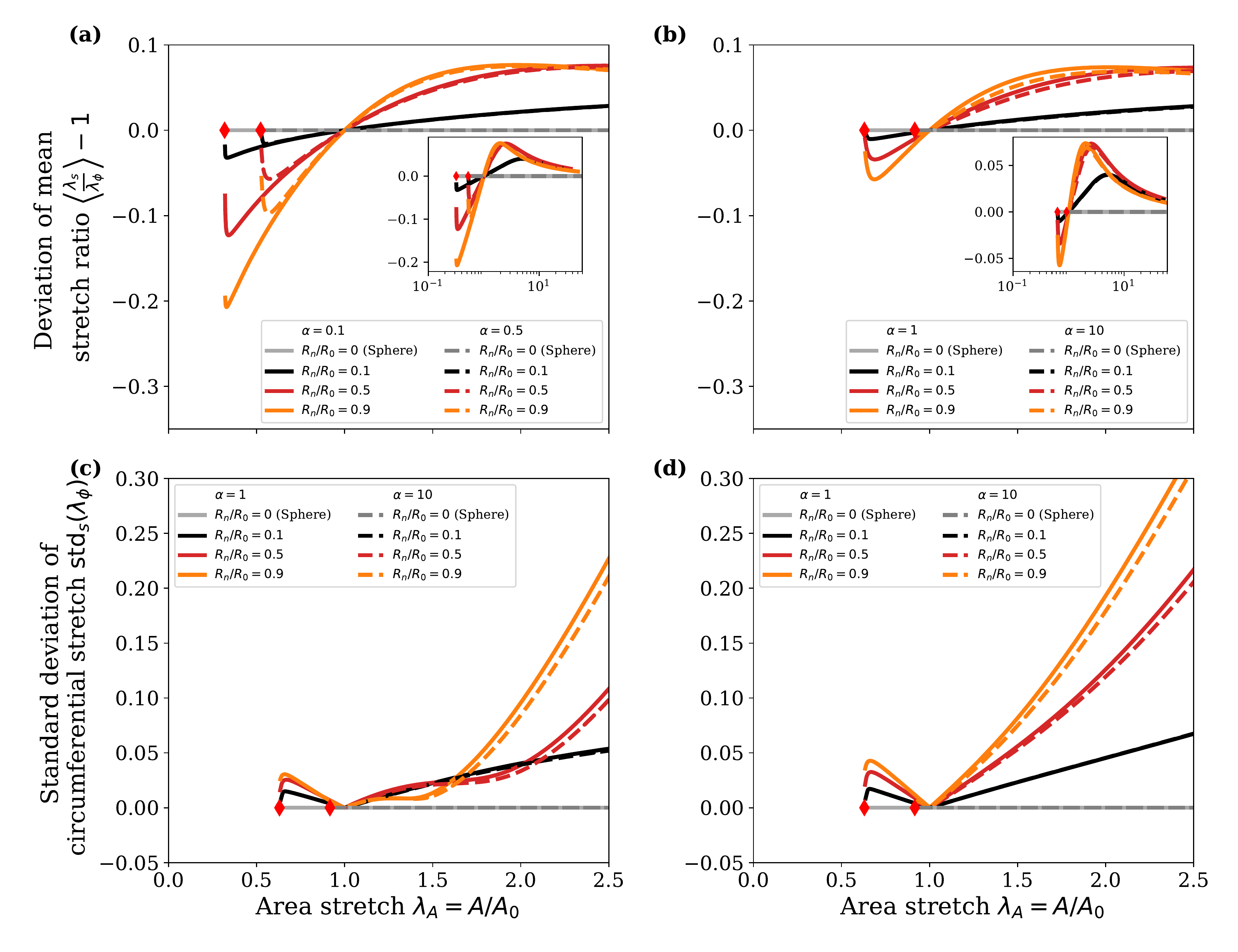}
    \caption{ Characterization of the stretch anisotropy  and the
      stretch inhomogeneity. (a,b) The 
      mean ratio of meridional and circumferential stretches
      $\left\langle \frac{\lambda_s}{\lambda_{\phi}} \right\rangle-1$
      along the contour characterizes stretch
       anisotropy and is shown 
       for  (a) $\alpha \leq 0.5$ and (b) $\alpha > 0.5 $.
       The standard deviations of  (c) meridional stretches
       $\lambda_s$ and (d) circumferential stretches $\lambda_{\phi}$
       along the contour characterize the inhomogeneity of
       stretches. We show the critical stretches $\lambda_{A,c}$ as red diamonds in (a-d).
  }
    \label{fig:anisotropy_area_stretch}
\end{figure*}

To investigate why the spherical sector approximations fit  the results so well, Fig.\ \ref{fig:anisotropy_area_stretch} plots different measures of the anisotropy of the stretch distributions on the droploon surface obtained from the Neo-Hookean shape equations for the same parameter ranges as in Fig.\ \ref{fig:PressureDeformationSPhereNeedle}. In the case of fully isotropic deformation, corresponding to a spherical sector shape, the deviation of the mean stretch ratio along the contour $\left\langle \frac{\lambda_s}{\lambda_{\phi}} \right\rangle-1$ (Fig.\ \ref{fig:anisotropy_area_stretch}a,b) and the standard deviation of the meridional and circumferential stretches $\mathrm{std}_s(\lambda_{s,\phi})$ (Fig.\ \ref{fig:anisotropy_area_stretch}c,d) are both zero. Since we neglect gravitational effects, it is clear that the unstressed shape
of the capsule at $\lambda_A = 1$ \emph{must} be a spherical sector. The
stretched shape will be anisotropically stressed, in general, because of the
boundary condition imposed by the attachment at the capillary.  We can find,
however, another particular stretch, where the \emph{stressed} shape is a
spherical sector.  This is reached at the critical stretch $\lambda_{A,c}$
(see also Section \ref{sec:ResultsSpheres}) at which $\Delta \hat {P}=0$.  The
force balance for every point on the capsule requires that the pressure force
cancels the tension force. For $\Delta \hat {P} = 0$, we therefore have
$\sigma_s = \sigma_\phi = 0$ all over the surface, i.e. the surface is
stress-free everywhere at this critical stretch. Since
$\sigma_s = \sigma_\phi = 0$ implies isotropic stretching, the shape at this
point is again correctly described by the spherical sector equation
(\ref{eq:PressureDefNeedle}).  If the stretch is further decreased to
$\lambda_A<\lambda_{A,c}$ both $\sigma_s<0$ and $\sigma_\phi<0$ will become compressive and buckling or  wrinkling instabilities of the droploon interface will 
occur \cite{LandauLifshitz,Knoche2013}.  

For stretch values other than $\lambda_A = 1$ or $\lambda_{A,c}$, the droploon
shape is non-spherical, because of the anisotropy
($\lambda_s \neq \lambda_\phi$) introduced by the boundary condition at the
capillary. This can clearly be seen in Figs.\
\ref{fig:anisotropy_area_stretch}a,b. For inflated shapes $\lambda_A > 1$, we find
$\left\langle \frac{\lambda_s}{\lambda_{\phi}} \right\rangle - 1 >0$
indicating that stretching is biased towards meridional deformations resulting
in slightly prolate shapes, whereas for deflated shapes $\lambda_A < 1$,
$\left\langle \frac{\lambda_s}{\lambda_{\phi}} \right\rangle-1<0$ and
circumferential deformations are preferred, resulting in slightly oblate
shapes.
 The mean anisotropy increases upon inflation before
decreasing again at much higher stretches (see the insets in
Fig.\ \ref{fig:anisotropy_area_stretch}a,b for a wider deformation range), when the influence of the capillary
becomes again negligible.

Furthermore, the standard deviation of the stretches along the contour
$\mathrm{std}_s(\lambda_s )$ and $\mathrm{std}_s(\lambda_\phi)$ shown in
Figs.\ \ref{fig:anisotropy_area_stretch}c,d characterizes the inhomogeneity of
the stretches along a contour. A standard deviation of
$\mathrm{std}_s(\lambda_s ) = \mathrm{std}_s(\lambda_s ) = 0$ corresponds to a
spherical sector. The meridional and circumferential stretches of an inflated
droploon are isotropic at the apex with
$\lambda_s(s = 0) = \lambda_\phi(s = 0)\propto \lambda_A^{1/2}$. At the
capillary, the attachment condition mandates $\lambda_\phi^\mathrm{cap} = 1$
while $\lambda_s^\mathrm{cap}$ increases with $\lambda_A$, which introduces
anisotropy and inhomogeneity into the problem with meridional stresses
accumulating at the capillary. The spherical approximation will hold well for
shapes where the stretches are approximately \textit{homogeneous} over a large
arc length, corresponding to a small standard deviation of the stretches, and
\textit{isotropic}, corresponding to a mean stretch along the contour
$\left\langle \frac{\lambda_s}{\lambda_{\phi}} \right\rangle$ close to unity.
This is fulfilled at the two spherical configurations
$\lambda_A = 1$ and $\lambda_{A,c}$. The spherical configuration 
with  $\lambda_{A,c}$ appears to be  highly sensitive, and small changes in
$\lambda_A$ lead to large deviations in the anisotropy (and inhomogeneity).
It is interesting to note that at small deformations around $\lambda_A= 1$,
the anisotropy evolution depends only on the ratio $R_n / R_0$ and not on $\alpha$.

We argue that the evolution of the anisotropy and inhomogeneity can be grasped by considering that the capillary acts similarly to a rigid inclusion
in a stretched elastic membrane as both enforce the absence of circumferential
stretching ($\lambda_\phi = 1$)
at their boundary.  A rigid inclusion
in a stretched elastic membrane is known to concentrate the meriodional
stresses creating anisotropy and inhomogeneity, similar to the stress
concentration around a crack tip.  For flat membranes, a rigid inclusion is a
classic problem that was studied for neo-Hookean membranes by Wong and Shield
\cite{Wong1969}. For the droploon we have a curved geometry, which gives rise to an even more pronounced  increase of  anisotropy around the capillary.

\begin{figure*}
    \centering
    \includegraphics[width=15cm]{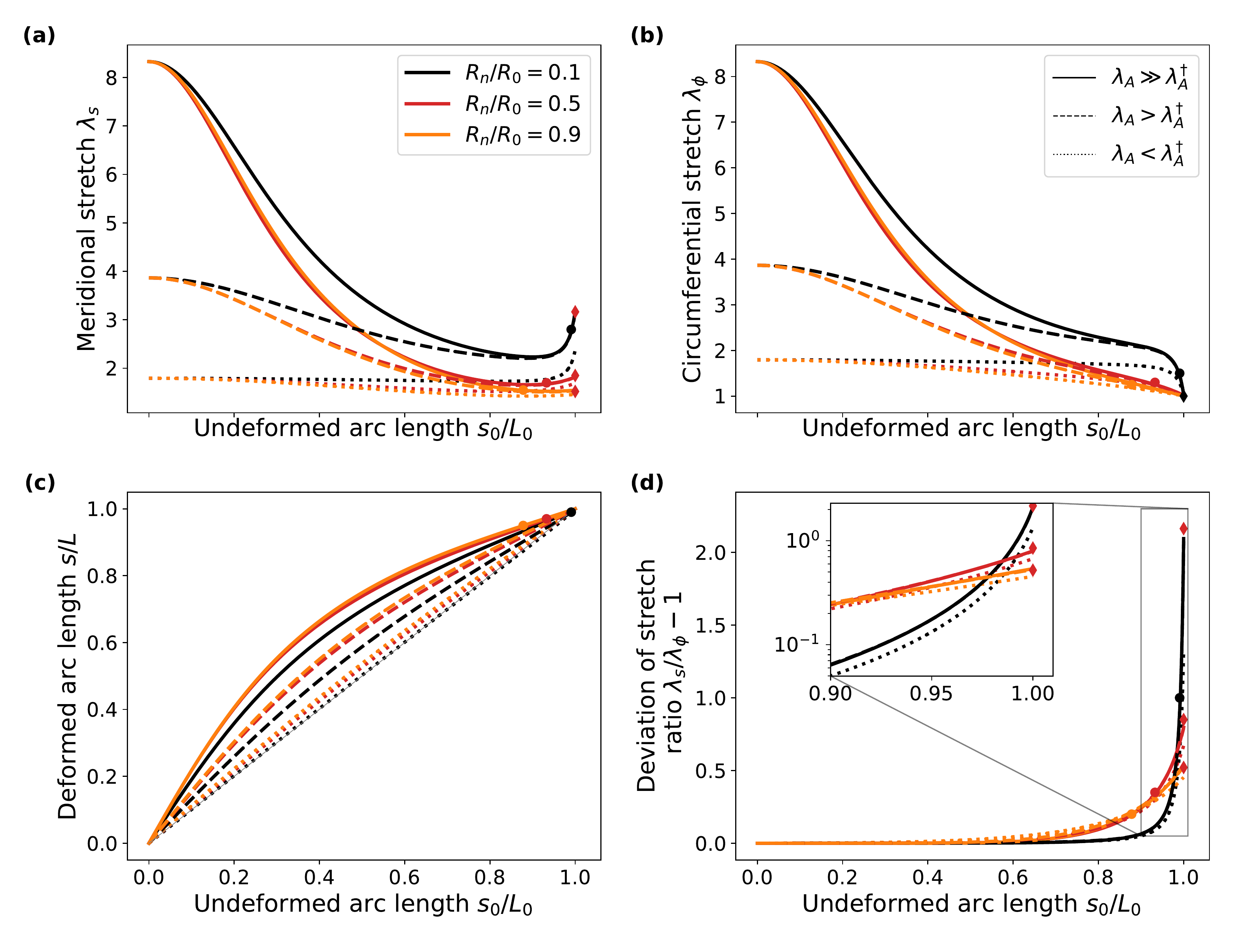}
    \caption{
      Stretch anisotropy of droploon shapes with $\alpha = 10$ for
      three values of $R_n/R_0$ for each of three area stretches
      $\lambda_A \gg \lambda_A^\dag$, $\lambda_A > \lambda_A^\dag$, and
      $\lambda_A < \lambda_A^\dag$ (see also Fig.\ \ref{fig:Q}
      for a definition of the characteristic 
      area stretch $\lambda_A^\dag$). 
      (a,b) Stretch ratios $\lambda_s$ and
      $\lambda_\phi$ as a function of the undeformed arc length $s_0/L_0$
      along the contour. While $\lambda_\phi$ is approaching the undeformed
      value of 1 at the capillary ($s_0/L_0=1$), $\lambda_s$ rises at the
      capillary. (c) shows that the deformed arc length $s$ considerably
      deviates from the the undeformed arc length $s_0$ along the contour. (d)
      The resulting stretch anisotropy $\lambda_s / \lambda_\phi - 1$ is
      localized at the capillary. The size of the anisotropy zone around the
      capillary can be characterized by an exponential decay arc
      length $s_0^*$,
      which is calculated from the logarithmic derivative of
      $\lambda_s / \lambda_\phi - 1$ at the capillary for the solid lines and
      shown as colored dots in all plots (a-d). We also show the maximal stretch at the capillary from Eq.\ (\ref{eqn:maximal_anisotropy}) as red diamonds in (a) and (d).  }
    \label{fig:decayDetails}
\end{figure*}

We see clear evidence of  the increased anisotropy 
around the capillary in  numerical solutions to
the full anisotropic shape equations from  Section
\ref{sec:ModelKierfeld} as shown in Fig.\ \ref{fig:decayDetails}. 
In 
Fig.\ \ref{fig:decayDetails}a,b,c,  we show the stretch ratios $\lambda_s$ and $\lambda_\phi$ and 
the redistribution of arc length along the contour of inflated 
droploons. These results show  the rise of meridional stretch 
close to the capillary.
Fig.\ \ref{fig:decayDetails}d reveals that the
resulting 
stretch anisotropy $\lambda_s/\lambda_\phi-1$ is  localized  at the
capillary and that it decays exponentially
over a characteristic arc length
$s_0^*$ away from the capillary.
Here, $s_0$ is the arc length of the undeformed reference shape
(the spherical droplet),
which is related to the arc length $s$ of the deformed
shape by the meridional stretch ratio,
$\mathrm{d}s / \mathrm{d}s_0 = \lambda_s$
(see section \ref{sec:ModelKierfeld}).
We use the logarithmic derivative
of  $\lambda_s/\lambda_\phi-1$ to numerically determine  the size $s_0^*$
of the zone of increased anisotropy around the capillary.

We propose that the relative meridional extent
 of the anisotropy zone  along
 the \emph{deformed} droploon contour provides a non-dimensonal number $Q$,
 which is suitable to 
 characterize the importance of elastic anisotropy effects
 in the regime $\alpha>1$, where  elastic energies dominate.
 We thus define $  Q \equiv {s^*}/{L} $, where
 $s^*$ is the meridional length  of the anisotropy region
 measured in terms of the \emph{deformed} arc length, while $L$ is the
 total arc length of the deformed droploon contour. 
 For $\alpha<1$,
 elastic energies are small compared to droplet surface tension such
 that also elastic anisotropy becomes less important. 

\begin{figure*}
    \centering
    \includegraphics[width=15cm]{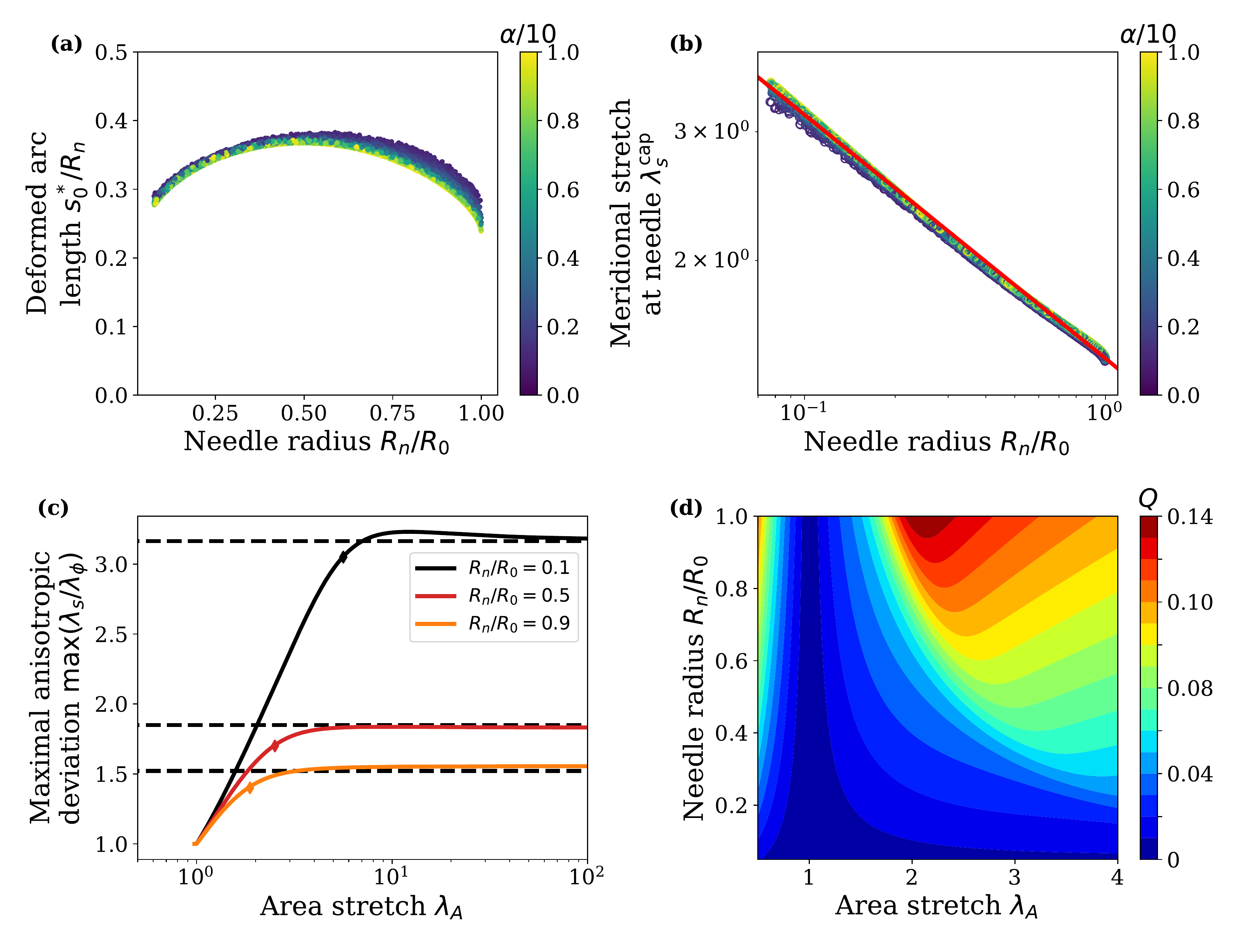}
    \caption{Analysis of the anisotropy zone and the anisotropy
        parameter $Q$ from
      numerical solutions of the anisotropic shape
      equations. (a) The size of the anisotropy zone  $s_0^*$ is roughly constant
      giving rise to the bound (\ref{eqn:decaylength}).
      (b) The saturation value is mainly determined by the
        parameter ($R_n/R_0$), see Eq.\
      (\ref{eqn:maximal_anisotropy}).
      (c)  As a function of the area stretch $\lambda_A$, the maximum
      anisotropy saturates at large deformations beyond a value
      $\lambda_A^\dag$ (results for $\alpha =10$ shown as colored diamonds). 
      (d) Contour plot of the  non-dimensional anisotropy parameter  $Q$
      according to Eq.\ (\ref{eqn:Q}).
      Stretch anisotropy effects are negligible for $Q\ll 1$.}
    \label{fig:Q}
\end{figure*}

In order to evaluate the anisotropy parameter $Q$, we 
use the general relation $\mathrm{d}s / \mathrm{d}s_0 = \lambda_s$
between deformed and undeformed arc length at the capillary
and $L  \sim \pi R_0 \lambda_A^{1/2}$ for the total arc length $L$
in the limit $R_n \ll R_0$ to obtain
\begin{equation}
  Q \equiv \frac{s^*}{L} \sim \frac{s_0^* \lambda_s^\mathrm{cap}}{L}
  \sim \frac{s_0^* \lambda_s^\mathrm{cap}}{\pi R_0} \lambda_A^{-1/2}
  \label{eqn:Q}
\end{equation}
where  $\lambda_s^\mathrm{cap}$ is the meridional stretch at the
capillary. To make further progress, we derive  relations
for the size $s_0^*$ of the anisotropy zone and the
stretch ratio $\lambda_s^\mathrm{cap}$ at the capillary
 from numerical results shown in  Fig.\ \ref{fig:Q}.

Because the  
 maximal stretch anisotropy is found at the capillary and 
 $\lambda_{\phi}=1$ at the capillary, the meridional stretch
 at the capillary actually equals the maximal stretch anisotropy, 
$\max{\left(\frac{\lambda_s} {\lambda_\phi}\right)} =
\lambda_s^\mathrm{cap}$. While in the case of flat membranes the maximal
aniosotropy $\lambda_s^\mathrm{cap} \propto \lambda_s(s=\infty)$ is
proportional to the radial stretch at infinity \cite{Wong1969},
our numerical results  for curved droploons indicate 
that $\lambda_s^\mathrm{cap}$ first increases upon inflation $\lambda_A>1$ but
saturates for highly inflated droploons with
area stretches  $\lambda_A$ exceeding
a fairly well-defined value $\lambda_A^\dag$,
as shown in Fig.\ \ref{fig:Q}c for the
case of $\alpha = 10$.  Further numerical analysis of the
saturation value as performed in  Fig.\ \ref{fig:Q}b
allows us to quantify the saturation value as 
\begin{equation}
  \max{\left(\frac{\lambda_s} {\lambda_\phi}\right)}
  \approx \lambda_s^\mathrm{cap} \equiv
  {\rm const} \left( \frac{R_n}{R_0}\right)^{-1/3} 
    \label{eqn:maximal_anisotropy}
  \end{equation}
  with ${\rm const} \approx 1.47$ in the regime $\alpha >1$. This saturation
  value is solely determined by the geometrical parameter $R_n / R_0$ of the
  undeformed droploon, which demonstrates that saturation is induced by
  droplet curvature.
  We also find $\lambda_A^\dag \sim (\lambda_s^\mathrm{cap})^{3 / 2}$
  for the
  area stretch, where saturation of the maximal anisotropy sets in. 
  The maximal anisotropy given in Eq.\ (\ref{eqn:maximal_anisotropy})
  diverges in the limit $R_n / R_0 \approx 0$, which
seems counter-intuitive at first, because the spherical approximation works
best for exactly this limit. This issue will be resolved below.

Let us quantify  the
size $s_0^*$ of the anisotropy zone around the capillary. From Fig.\ \ref{fig:Q}a, we find a conservative bound
\begin{equation}
   s_0^* \leq \frac{R_n}{2}.
    \label{eqn:decaylength}
  \end{equation}
This relation reveals that the size of the
stretch anisotropy zone
 is set by the geometry parameter $R_n/R_0$ of the reference
 state rather than  the elastocapillary number $\alpha$.

 Using Eq.\ (\ref{eqn:decaylength}) for $s_0^*$
 and the saturation value given in Eq.\ (\ref{eqn:maximal_anisotropy})
 for $\lambda_s^\mathrm{cap}$ in (\ref{eqn:Q}), we obtain 
\begin{equation}
  Q = \frac{\rm const}{2\pi}  \left(\frac{R_n}{R_0}\right)^{2 / 3}
   \frac{1}{\lambda_A^{1 / 2}} 
   \label{eqn:QLargeLambda}
 \end{equation}
 for the anisotropy parameter $Q$ for highly inflated droploons
 $\lambda_A>\lambda_A^\dag$.
 This parameter remains small  for $R_n \ll R_0$ indicating that
 we can neglect anisotropy effects in this limit.

 At smaller deformations $1 < \lambda_A < \lambda_A^\dag$, 
 where
 saturation of the capillary anisotropy has not yet set in, we numerically find
 that the maximal stretch anisotropy scales
 with $\log(\lambda_A)$ (see Fig.\ \ref{fig:Q}c), giving
\begin{equation}
  Q = \frac{R_n}{R_0} \frac{\lambda_s^\mathrm{cap} - 1}
  {3 \pi \log(\rm \lambda_s^\mathrm{cap})}
  \frac{\log(\lambda_A)}{\lambda_A^{1 / 2}},
    \label{eqn:anisotropyQuantifier}
\end{equation}
where we again use the saturation value $\lambda_s^\mathrm{cap}$ from Eq.\ (\ref{eqn:maximal_anisotropy}).

We obtain a full contour plot of the anisotropy
parameter $Q$ in Fig.\ \ref{fig:Q}d by
joining the results in the two regimes
( $\lambda_A > \lambda_A^\dag$ and $\lambda_A < \lambda_A^\dag$) with
a smooth interpolating function. This plot confirms that $Q$
is small ($Q \ll 1$) for shapes where the spherical approximation works
best. In particular, we find that
we can neglect anisotropy effects ($Q\ll 1$)
in the limit $R_n / R_0 \approx 0$,
resolving the counter-intuitive behaviour of the maximal anisotropy.  We
emphasize the fact that Eq.\ \eqref{eqn:anisotropyQuantifier} only
depends on
$R_n / R_0$ and $\lambda_A$ and \emph{not} on $\alpha$, as long as
$\alpha > 1$. This indicates  that stretch anisotropy is
mainly governed by geometry rather than by elastic energy
contributions.
As already pointed out above, elastic contributions and, thus,
also elastic anisotropy effects become increasingly irrelevant
for  $\alpha < 1$, where surface tension dominates and the
shape resembles a  spherical liquid droplet.
The regions $\lambda_A > \lambda_A^\dag$ and $\lambda_A < \lambda_A^\dag$
differ markedly in their functional dependence on 
 $\lambda_A$.
This results in a maximum of the parameter $Q$ for area stretches
$\lambda_A\sim \lambda_A^\dag\propto (R_n / R_0)^{- 1 / 2}$
at a fixed value of $R_n/R_0$.
This, in turn, indicates that stretch anisotropy is most relevant
for these intermediate area stretches. 

\begin{figure*}
\centering
       \includegraphics[width=\linewidth,keepaspectratio]{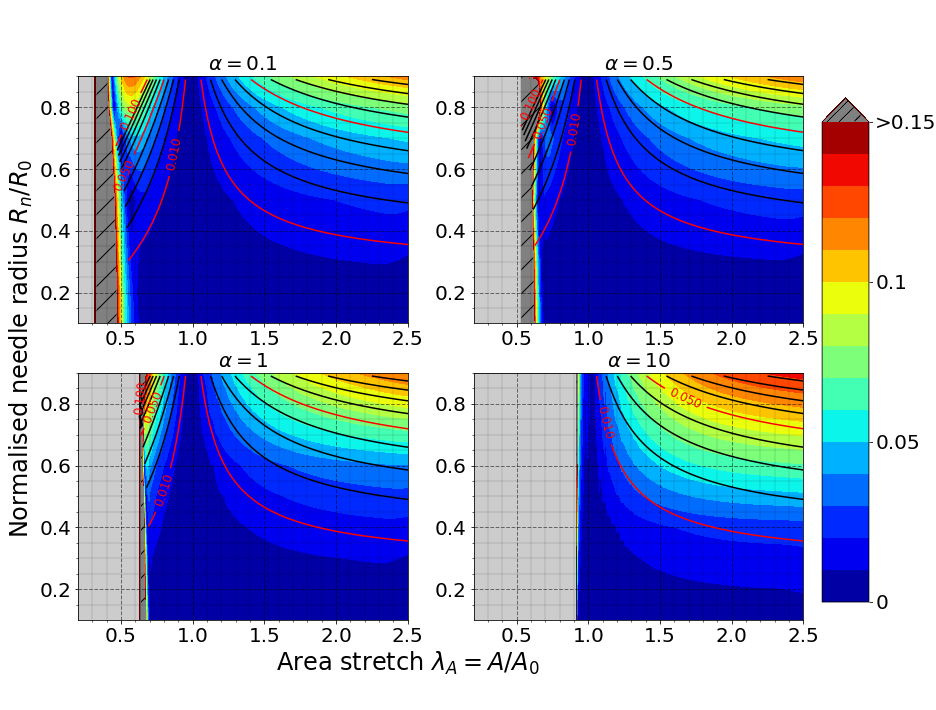}
       \caption{Relative error of pressure difference between Surface Evovler and neo-Hookean perfect sphere, at the same area stretch $\lambda_A$ for four elastocapillary numbers ($\alpha=0.1$,$0.5$,$1$,$10$). The grey boxes delimit the stretch values below critical stretch value $\lambda_{A,c}$. Full lines are lines of equal relative error between the neo-Hookean perfect sphere and the neo-Hookean truncated sphere, given by Equ. \eqref{eq:ErrorHeatMapSpherical}. }
     \label{fig:ErrorDroploonNeedle}
\end{figure*}

The possibility of approximating the droploon shape by a spherical sector over a wide range of parameters is an important piece of information for experimentalists  since it means that the analytical expression of Eq. (\ref{eq:PressureDefNeedle}) can be used to quantify reliably the elastocapillary properties of the droploon interfaces over a reasonably wide range of elastocapillary numbers. We also remind the reader that from the expressions it evident that within our geometrical approximations, the critical area stretch at which the pressure changes sign is independent of the size of the capillary. 
The combined numerical analysis provides another important piece of information: for reasonably small capillary sizes ($R_n/R_0<0.5$), the pressure-deformation relation is actually well described by the simple sphere equations without capillary (Section \ref{sec:TheorySpheres}), making the quantitative interpretation of experimental data fairly straightforward. In order to quantify the deviation from the simple sphere theory, we plot  in Fig.\ref{fig:ErrorDroploonNeedle} the heatmap of the normalised deviation of the numerically predicted pressure $\Delta\hat{P}$ with capillary (using Surface Evolver) from that predicted by the sphere theory $\Delta \hat{P}_S$ for a given area stretch, i.e. we plot
\begin{equation}
    \left| \frac{\Delta\hat{P}_S-\Delta\hat{P}}{\Delta\hat{P}}\right| = \left| 1 - \frac{\Delta\hat{P}_S}{\Delta\hat{P}}\right|.
    \label{eq:ErrorHeatMap}
\end{equation}
Making the spherical sector hypothesis of Eq. (\ref{eq:PressureDefNeedle}), this expression becomes simply 
\begin{equation}
    \left| 1 - \frac{\Delta\hat{P}_S}{\Delta\hat{P}} \right| = \left| 1 - \mathpzc{f} \right|,
    \label{eq:ErrorHeatMapSpherical}
\end{equation}
which is plotted as lines of equal relative error. These isolines are identical in all four graphs of Fig.\ref{fig:ErrorDroploonNeedle} since they are independent of $\alpha$ (see Eq.\eqref{eq:GeometricalCorrection}). 

Deviations of the heatmaps in Fig. \ref{fig:ErrorDroploonNeedle} from the geometrical prediction have two origins: imperfect relaxation in the simulations and the influence of shear contributions of the solid skin which are neglected in the geometrical approximations. The first is at the origin of most of the deviations for $\alpha < 10$, while the latter starts to be clearly visible for $\alpha = 10$. Nevertheless, this latter difference remains small ($<0.5\%$), confirming again that shear contributions play a minor role in most of the investigated parameter range in accordance with the non-dimensional $Q$-parameter plotted in Fig. \ref{fig:Q}d. Our geometrically-corrected pressure-deformation relation of Eq. (\ref{eq:PressureDefNeedle}), although not accounting for stretch anisotropy, is therefore a very good approximation for pendant drops with Neo-Hookean elastic interfaces within the parameter range investigated here. 

Let us now turn to the analysis of the heatmaps themselves. They indicate that in the small deformation limit ($\lambda_A \approx 1$), the error made in using the sphere approximation remains smaller than $1\%$ at any radii ratio and elastocapillary number. For larger deformations in the inflation regime ($\lambda_A>1$), the approximation error is still smaller than $1\%$ for small capillary radii ($R_n/R_0<0.2$). Similar behaviour is observed in the deflation regime. However, the prediction systematically fails when approaching the critical stretch $\lambda_{A,c}$. This is because wrinkling instabilities in the skin may become relevant in this regime. This phenomenon can be captured neither within the sphere approximation, nor by our Surface Evolver simulations where the skin bending energy - crucial for wrinkling - is not taken into account. Skin bending can be implemented in Surface Evolver, but is beyond the scope of this paper. In the heatmaps we have therefore colored these zones in  gray. 

At small $\alpha$ and large $R_n/R_0$ an additional  zone of large approximation error ($>10\%$) appears for pressures $\Delta\hat{P}\approx 1$. This deviation arises from the increasing difference between sphere and truncated sphere geometry: As the truncated sphere shrinks, it reaches the shape of a half-sphere of radius $R_n$. Any further decrease in drop volume causes an actual increase in curvature radius which is not captured by the sphere theory, hence the failure of the analytical prediction beyond this point in the parameters space.

Despite those considerations for large capillary radii, the heatmaps of Fig.\ref{fig:ErrorDroploonNeedle} provide very good news for the experimentalist aiming to quantify the elastic properties of droploon surfaces: when working with reasonable capillary sizes ($R_n/R_0<0.5$), reasonably small deformations (<0.1) and reasonable elastocapillary numbers ($\alpha < 10$), experimental data can be confidently fitted by the simple sphere theory (without capillary) since experimental errors are likely to outweigh the small error introduced by the sphere assumption.


\clearpage

\newpage

\section{Conclusion and outlook}

Treating the seemingly simple problem of a drop covered by an elastic skin attached to a circular capillary in the absence of gravity, we have been able to show that Surface Evolver simulations are a powerful tool to study systems in which surface tension and nonlinear (Neo-Hookean) elasticity co-exist within the same interface. We have chosen on purpose such a simple geometry, in order to avail of independent theoretical and numerical predictions relying on cylindrical symmetry (Section \ref{sec:Theory} and Section \ref{sec:ModelKierfeld}) which can be compared to the Surface Evolver solutions. In all cases, they showed excellent agreement. Surface Evolver will therefore be useful to tackle more complex geometries, such as droploons on complex capillary shapes, interacting droploons or complete emulsions composed of droploons, where theory or alternative numerical predictions requiring symmetry will not be available. In contrast to other finite element tools, the energy minimisation approach of Surface Evolver, widely used in the communities studying foams and emulsions, provides access to a wide range of problems in which   interfaces of complex geometry play a key role. In the Appendix we provide a detailed description of the implementation of nonlinear elasticity in Surface Evolver simulations   to facilitate future developments, and we also provide our Surface Evolver code for download in the Supplementary Materials. Taking into account bending stiffness in the simulations would be an interesting perspective for future work.

For simplicity, we have been talking about drops/droploons all along. However, all presented concepts are equally valid for bubbles/bubbloons and hence foams. Our analysis shows how complex the interplay of capillary and elastic forces at an interface is, even for the relatively simple geometry of an initially spherical droploon inflated on a circular capillary.  Due to the intricate coupling of changes in interfacial curvature and area, accurate theoretical models and simulations are required  to extract interfacial properties quantitatively from measured pressure-deformation relations.

The problem of the pressure deformation of a
droploon covered by an elastic skin and attached to a capillary in the absence of gravity is a seemingly simple problem. From the point of view of elasticity theory it is challenging, however, because the elastic skin represents a closed curved shell and the capillary
a rigid circular inclusion within this shell.
Holes or rigid inclusions in elastic membranes are known to produce stress anisotropies and stress concentration upon stretching. Here, the droploon skin is stretched by inflation, contains a rigid inclusion and features the additional  complication of a background curvature because the initial relaxed shape is spherical
(neglecting gravity). We obtained theoretical predictions regarding the influence of the stress anisotropy induced by the capillary
onto the  pressure-deformation relation from Surface Evolver simulations and a careful numerical analysis
of stresses and strains in the shape equation approach. 
A full analytical solution remains an open problem for future research.

In the parameter range investigated by our simulations, we have been able to show that for elastocapillary numbers of $\alpha < 10$ the influence of the capillary on the pressure-deformation relation is essentially of geometrical nature, i.e. the capillary modifies in the first place the relationship between the area stretch (related to interfacial stress) and the interface curvature. In this case, the droploon shapes can be represented approximately by spherical sectors and the pressure-deformation relation is given by Eq. (\ref{eq:PressureDefNeedle}). For interfaces with Gibbs elasticity, this expression is exact, while for (Neo-)Hookean interfaces it remains an (excellent) approximation. Deviations from this simple geometrical approximation are starting to be significant for the largest capillary sizes ($R_n/R_0=0.9$) and elastocapillary number ($\alpha = 10$) simulated by us, suggesting that the anisotropic contribution to the interfacial stress and deformation near the capillary is starting to play a role. 

To show that  this anisotropy is indeed strongly localised at the capillary, we calculate, as a function of position on the interface, the deviation of the ratio of meriodional and circumferential stretches from one. This quantity decays nearly exponentially with the distance  from the capillary, over a characteristic length $s^*$. The   extent of this anisotropically strained zone can be compared to the total droploon size by defining the non-dimensional ratio $Q=s^*/L$,  where $L$ is the total arc length of the droploon. For droploon inflation and for $\alpha > 1$, we find  
\begin{equation}
  Q = \frac{R_n}{R_0} \frac{\rm \lambda_s^\mathrm{cap} - 1}{3 \pi \log(\rm \lambda_s^\mathrm{cap})} \frac{\log(\lambda_A)}{\lambda_A^{1 / 2}},
    \label{eqn:anisotropyQuantifier2}
\end{equation}
with 
\begin{equation}
  \lambda_s^\mathrm{cap} \equiv
  {\rm const} \left( \frac{R_n}{R_0}\right)^{- 1 / 3}, 
    \label{eqn:maximal_anisotropy2}
  \end{equation}
being the "saturation" meridional stretch reached at the capillary for large deformations and const = 1.47. 
For large deformations, we therefore obtain
\begin{equation}
   Q =  \frac{\rm const}{2\pi} \left(\frac{R_n}{R_0}\right)^{2/3} \frac{1}{\lambda_A^{1 / 2}}. 
   \label{eqn:QLargeLambda2}
\end{equation}
These relations and their analysis provided in Section \ref{sec:ResultsDropsNeedles} and Fig. \ref{fig:Q} put in evidence that the extent of the anisotropic zone (and hence its influence on the pressure-deformation relation), is mainly controlled by the reference geometry of the droploon ($R_n/R_0$) and by the stretch $\lambda_A$. We therefore show for the first time that the extent of this zone is essentially governed by geometrical features while the influence of the elastocapillary number $\alpha$ remains negligible. These are very good news for experimentalists who can rely on the spherical droploon equations given in Table \ref{tab:models} combined with the geometrical correction of Eq. (\ref{eq:GeometricalCorrection}) to fit their data for a wide range of $\alpha$  as long as $R_n/R_0$ and $\lambda_A$ remain reasonable. The heatmaps and relations provided in Section \ref{sec:ResultsDropsNeedles} will help to estimate the appropriate parameter ranges.

More importantly for the analysis of experimental data, we have also shown that when working with sufficiently small capillaries ($R_n/R_0<0.5$) and at small deformations ($\sim 5\%$ area), the simple analytical pressure-deformation relations of spheres \textit{without} capillaries (Table \ref{tab:models}) provide excellent approximations to the pressure-deformation relations of droploons on capillaries. The much simpler analytical relations of Table \ref{tab:models} can therefore be used to extract quantitative interfacial properties from fits to  experimental data. Experimentalists are referred to Fig. \ref{fig:ErrorDroploonNeedle} to estimate the error they make using this approximation. 

In Section \ref{sec:TheorySpheres} we showed that  for small deformations, the Gibbs, Neo-Hookean and Hookean models for liquid- and solid-like interfaces all predict the same kind of pressure-deformation relation. In view of the analysis presented above, this may explain why a lot of experimental data for solid-like interfaces seems to have been successfully fitted in the past by the Gibbs model. Indeed, our analysis shows that at small deformations, pendant drop experiments with nearly spherical droploons do not allow to discriminate between liquid-like and solid-like interfaces.   Alternative experiments, such as interfacial shear rheology measurements or the Capillary Mensicus Dynanometry\cite{Danov_CollIntSci_2015} are required to obtain this information. 

We have chosen here a minimal model of a droploon interface where the elastic extra stress of a Neo-Hookean solid material is simply added to a constant interfacial tension. Real interfaces are not as simple \cite{Edwards1991,Rehage_RheoAct_2002,Sagis_RevModPhys_2011,Erni2011,Fuller_SoftMatter_2011,Fuller_2012,Sagis_COCIS_2014,Verwijlen_ACIS_2014,Pepicelli_SocRheo_2019}. Surface tension and elasticity tend to be coupled in a complex manner \cite{Verwijlen_ACIS_2014}, and the description of the response of the elastic membrane is likely to require taking into account an anisotropic, viscous and plastic  response as well as non-linearities which are more complex than those of the Neo-Hookean model. Nevertheless, our simple approach  already gives important insight into some fundamental properties of pressure-deformation relations of pendant droploons.

Considering that pendant drop experiments, even in the simplest configuration without gravity, overlay a geometric non-linearity with non-linearities in the material response of a solid-like interfacial material, it remains  questionable if this is the appropriate experimental choice to discriminate between appropriate models to describe solid-like interfaces. Differences between models are likely to show up only at larger deformations which makes the interpretation extremely difficult. However, due to their simplicity, pendant drop experiments remain an excellent choice for a phenomenological characterisation of the dilational visco-elastic properties at small deformation.

Last but not least, all our investigations have been performed without gravity, while pendant drops (and bubbles) are prone to gravity-driven deformations rendering them non-spherical. We recall that for a nearly spherical drop the Bond number $ Bo = \Delta\rho g R_0^2/\gamma_0$ indicates the ratio of the hydrostatic pressure difference between the top and the bottom of the bubble $\Delta\rho g 2R_0$ and the Laplace pressure which is due to surface tension $2 \gamma_0 /R_0$. The impact of gravity on bubble shape is negligible if $Bo \ll 1$. If density-matched systems cannot be used, very small bubbles may therefore be a solution \cite{Kotula_JRheo_2015} to reduce the impact of gravity. This also has the advantage to increase the interface curvature, and hence the pressure and therefore experimental sensitivity.

If gravity-driven deformation cannot be completely avoided, the following two aspects need to be taken into account. The first influence of gravity is on the shape of the droploon in the reference state. Gravity may create a concave neck close to the capillary, which creates additional stress localisation. Using  numerical investigations of the droplet shape bifurcation diagram (yellow line of bifurcations in Figs.\ 4 and 5 of Ref.\ \cite{Kratz2020}), we could show in previous work that only for
\begin{equation}
\frac{R_n}{R_0} < 2.6 \mathrm{Bo}^{1.64},
\end{equation}
the drop remains fully convex and neck formation can be neglected. 

The second aspect concerns deformation with elastic skins, where the increasing droploon size upon inflation or the decreasing effective surface stresses upon deflation may make the system increasingly sensitive to gravity. In this case one may want to introduce an elastic Bond number which contains the deformation-dependent elastic contribution to the surface stress based on the Hookean expression  (\ref{eq:HookeTension})

\begin{equation}
    Bo_{el} = \frac{\Delta \rho g}{\gamma_0(1+2\alpha(\lambda-1))}\lambda^2 R_0^2.
\end{equation}

For sufficiently small elastic Bond numbers, gravity can then be neglected. Since gravity can be implemented easily in Surface Evolver, future investigations may explore the influence of gravity more quantitatively.

\section*{Acknowledgements}
This work has been conducted in the framework of an ERC Consolidator Grant (agreement 819511 – METAFOAM). It has also profited from an IdEx Unistra "Attractivity grant" (Chaire W. Drenckhan) and has, as such, benefited from funding from the state, managed by the French National Research Agency as part of the 'Investments for the future' program. The authors would like to thank François Schosseler, Leandro Jacomine, Stephane Pivard and Aurélier Hourlier-Fargette for regular in-depth discussions concerning the experimental characterisation of pressure-deformation relations of droploons, which has stimulated greatly this numerical investigation.

\section{Appendix A : numerical determination of the interfacial deformation}
\label{AppendixA}

We use the Surface Evolver software to determine the bubble or droplet shapes for which interfacial energy is minimal, respecting volume constraints and boundary conditions. The case where an elastic skin is attached to the interface raises the question how local strain should be deduced from the representation of the interface as an assembly of triangular facets. Section \ref{sec:convected} explains how {\it convected } coordinates are used for this. Section \ref{sec:energy} provides details about the calculation of the elastic energy density, based on the neo Hooke constitutive model. 
\subsection{Strain represented using convected coordinates \label{sec:convected}}
\begin{figure}
\centering
       \includegraphics[width=\linewidth,keepaspectratio]{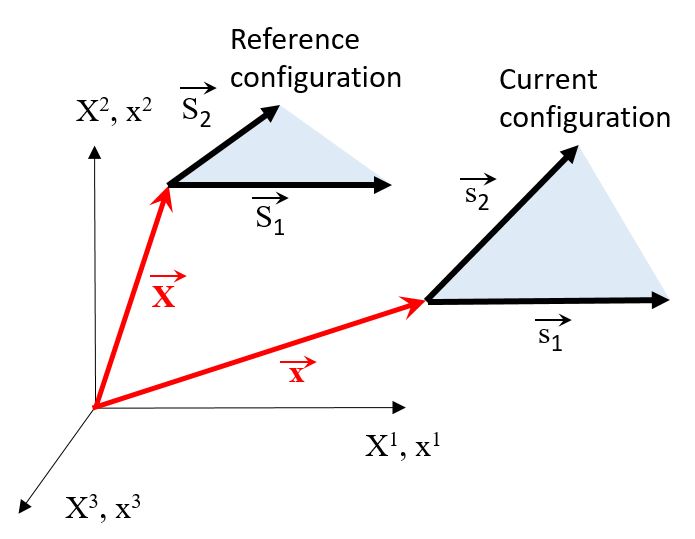}
       \caption{A triangular finite element of  an interface is represented in the reference configuration and in the current, deformed configuration. The figure illustrates the notations used in the text: $\vec{\bf x}$ for vectors pointing to vertices   and $\vec{\bf s}$ for finite element edge vectors. Capital letters  are used for the reference configuration and small letters for the current configuration. For the sake of simplicity, only one of the three vectors pointing to vertices is shown in each configuration. The contravariant components of both $\vec{\bf X}$ and $\vec{\bf x}$ are indicated on the same set of Cartesian axes.}
     \label{fig:convected}
\end{figure}
The shape of the triangular facets used in the Surface Evolver as finite elements is fully defined if two edge vectors are given. Upon deformation of the investigated bubble, the facet is generally displaced and the edge vectors are changed, spanning a facet of modified shape.  In the spirit of a linear discretization, an affine displacement field is assumed within each facet. One could describe the facet deformation using a coordinate system  whose origin is attached to a given vertex of the facet, and express how the Cartesian coordinates of each point on the facet evolve.  Alternatively, one may interpret the edge vectors as  basis vectors which evolve upon a deformation  and which are therefore in general non orthogonal. In this latter approach, the coordinates of each point of the interface are fixed and  the deformation is   represented in terms of a change of the basis vectors. This "convected coordinate" method goes back to pioneering work by Hencky \cite{Hencky1925}.
In the Surface Evolver this method is convenient because the relevant edge vectors can easily be derived from the three facet vertex positions in the current configuration, denoted  $\vec{x}_1,\vec{x}_2,\vec{x}_3$ and in the reference configuration $\vec{X}_1,\vec{X}_2,\vec{X}_3$,
\begin{equation}
\label{eq:defS}
\begin{split}
      \vec{S}_1&=\vec{X}_3-\vec{X}_1 , \quad  \vec{s}_1=\vec{x}_3-\vec{x}_1,\\
    \vec{S}_2&=\vec{X}_2-\vec{X}_1 ,   \quad  \vec{s}_2=\vec{x}_2-\vec{x}_1.
\end{split}
\end{equation}
The edge vectors are represented using a cartesian orthonormal basis $(\vec{e}_x,\vec{e}_y)$ such that $\vec{S}_i=S_{ix}\vec{e}_x+S_{iy}\vec{e}_y$ and  $\vec{s}_i=s_{ix}\vec{e}_x+s_{iy}\vec{e}_y$. \par    
As mentioned, convected coordinates remain constant upon a deformation; this introduces simplicity. But this choice also introduces complexity since the expression of the scalar product is no longer given by contraction $\vec{a}\cdot\vec{b}=a_ib_i$, additional terms appear since the basis vectors are generally not orthogonal. To avoid such complexity, one represents vectors and tensors that one wishes to associate in products using two different bases: a "covariant" and contravariant one. Covariant basis vectors follow the deformation of the edge facets. They are denoted $\vec{G}_1,\vec{G}_2$ in the reference state and $\vec{g}_1,\vec{g}_2$ in the current state. Covariant quantities are identified by lower indices, 
\begin{equation}
\begin{split}
     \vec{G}_1&= \vec{S}_1,\quad  \vec{g}_1= \vec{s}_1,\\
    \vec{G}_2&=\vec{S}_2,\quad  \vec{g}_2= \vec{s}_2.
\end{split}
\end{equation}
Contravariant basis vectors $(\vec{G}^1,\vec{G}^2) $ or $(\vec{g}^1,\vec{g}^2 )$, are identified by upper indices, and they are defined through the following orthogonality relations:
\begin{equation}
    \vec{G}^i\cdot\vec{G}_j=\delta^i_j, \quad  \vec{g}^i\dot \quad\vec{g}_j=\delta^i_j,
    \label{eq:orthogonality}
\end{equation}
where  $\delta^i_j =1$ if $i=j$ and $\delta^i_j =0$  otherwise. The Cartesian coordinate system is a special case within this general framework where covariant and contravariant bases coincide. Using co- and contravariant bases simplifies the expressions of the scalar products of vectors and tensors in the case of non-orthogonal basis vectors.

An arbitrary vector $d\vec{X}$ representing a small line element on the surface reads in terms of the covariant basis
\begin{equation}
    d\vec{X} = d\Theta^j \vec{G}_j.
\end{equation} 
 $d\Theta^j$ are the convected  contravariant coordinates. We use the Einstein summation convention
 and sum over repeated indices.
 
Descriptions of strain in large deformation continuum mechanics are commonly based on the deformation gradient tensor $\mathbf{F}$, represented by a matrix that transforms  a line element $ d\vec{X} $ in the reference state into  $ d\vec{x} $ in the current state,
\begin{equation}
  d\vec{x} = \mathbf{F} d\vec{X}.  
\end{equation}
In terms of convected coordinates, $\mathbf{F}$ may be written
\begin{equation}
\mathbf{F}=g_j\otimes G^j.
\label{Equation:RightCauchyGreen}
\end{equation}
The symbol $\otimes$ indicates an operation assembling two vectors into a tensor, called tensor product. 
Indeed, in view of Eq.\ \ref{eq:orthogonality} we have
\begin{equation}
   \mathbf{F} d \vec{X} = (\vec{g}_i\otimes \vec{G}^i)\,d\Theta^i G_j=d\Theta^i \vec{g}_i = d \vec{x}.
\end{equation}
The deformation gradient tensor contains information about rotations that is irrelevant for interfacial energy. The interfacial energy in the Surface Evolver  is computed using the 2D right Cauchy-Green strain tensor $\mathbf{C}$ which is invariant to rotations\cite{Mal1991}, contrary to $\mathbf{F}$:
\begin{equation}
\mathbf{C}=\mathbf{F}^T \mathbf{F}=(\vec{G}^i\otimes \vec{g}_i)(\vec{g}_j\otimes \vec{G}^j)=g_{ij}\,\vec{G}^i\otimes \vec{G}^j.
\label{Equation:RightCauchyGreenAppendix}
\end{equation}
$g_{ij}$ is the metric tensor in the current configuration, defined as follows:
\begin{equation}
     g_{ij} =\vec{g}_i \cdot \vec{g}_j.
     \label{eq:metric}
\end{equation}

To determine the elastic energy of a facet in a simulation, {\bf C} needs to be determined numerically. The components of the contravariant   basis vectors in the reference state $G^i$ are deduced from the covariant ones using the   orthogonality properties (\ref{eq:orthogonality}): 

\begin{equation}
\begin{aligned}
\vec{G}^1\cdot\vec{G}_1=1=S^{1x}S_{1x}+S^{1y}S_{1y} &\rightarrow &S^{1x}=\frac{1-S^{1y}S_{1y}}{S_1x}\\
\vec{G}^2\cdot\vec{G}_1 = 0 =S^{2x}S_{1x}+S^{2y}S_{1y} & \rightarrow & S^{2x}=-S^{2y}\frac{S_{1y}}{S_{1x}}\\
\vec{G}^2\cdot\vec{G}_2 = 1 =S^{2x}S_{2x}+S^{2y}S_{2y} &  \rightarrow & S^{2x}=\frac{1-S^{2y}S_{2y}}{S_{2x}}\\
\vec{G}^1\cdot\vec{G}_2 = 0 =S^{1x}S_{2x}+S^{1y}S_{2y} & \rightarrow & S^{1x}=-S^{1y}\frac{S_{2y}}{S_{2x}}.
\end{aligned}
\label{Equation:OrthogonalMaterialBasis}
\end{equation} 

Solving the system (\ref{Equation:OrthogonalMaterialBasis}) yields the components of the vectors $\vec{G}^i$: 
\begin{equation}\begin{aligned}
\vec{G}^1=\vec{S}^1=\left(\frac{S_{2y}}{S_{1x}S_{2y}-S_{1y}S_{2x}}, -\frac{S_{2x}}{S_{1x}S_{2y}-S_{1y}S_{2x}}\right) \\
\vec{G}^2=\vec{S}^2=\left(-\frac{S_{1y}}{S_{1x}S_{2y}-S_{1y}S_{2x}}, \frac{S_{1x}}{S_{1x}S_{2y}-S_{1y}S_{2x}}\right).
\end{aligned}
\label{Equation:ContravariantMaterialCoordinates}
\end{equation}

To express the Cauchy Green strain tensor directly as a function of the edge vectors, it is convenient to introduce Gram matrices. The Gram matrix of two arbitrary vectors $\vec{v}_1$ and $\vec{v}_2$ is a 2x2 matrix whose element $ij$ is by definition given by the scalar product $\vec{v}_i\cdot \vec{v}_j$. The covariant metric tensor defined in Eq.\ (\ref{eq:metric}) is thus the Gram matrix of the edge vectors in the current configuration. Following  the notation used in the Surface Evolver manual,  we will call this quantity  $\mathbf{s}$: 
 \begin{equation}
 \mathbf{s} = \begin{pmatrix}
 \vec{s}_1\cdot\vec{s}_1 & \vec{s}_1\cdot\vec{s}_2 \\
 \vec{s}_2\cdot\vec{s}_1 & \vec{s}_2\cdot\vec{s}_2
 \end{pmatrix}=g_{ij}.
 \end{equation}

The Gram matrix of the edge vectors in the reference state is denoted $\mathbf{S}$:
\begin{equation}
 \mathbf{S} = \begin{pmatrix}
 \vec{S}_1\cdot\vec{S}_1 & \vec{S}_1\cdot\vec{S}_2 \\
 \vec{S}_2\cdot\vec{S}_1 & \vec{S}_2\cdot\vec{S}_2
 \end{pmatrix}.
 \end{equation}

We note that the denominators in Eqs.\ (\ref{Equation:ContravariantMaterialCoordinates}) are the determinant of $ \mathbf{S}$: 
\begin{equation}
\mathrm{det}\,
 \mathbf{S} = \left(\vec{S}_1\cdot\vec{S}_1\right)\cdot\left(\vec{S}_2\cdot\vec{S}_2\right) - \left(\vec{S}_1\cdot\vec{S}_2\right)^2=\left(S_{1x}S_{2y}-S_{1y}S_{2x} \right)^2.
\label{Equation:CovariantGramDeterminant}
\end{equation}

Since the components of the tensor $G^i\otimes G^j$ are the scalar products of  $G^i$ and $G^j$ \cite{Kelly} we can now write Eq.\ (\ref{Equation:RightCauchyGreenAppendix}) in terms of the cartesian components of S, using Eqs.\ (\ref{Equation:ContravariantMaterialCoordinates}) and (\ref{Equation:CovariantGramDeterminant}),

\begin{equation}\begin{aligned}
\vec{G}^1\cdot\vec{G}^1 &= \frac{S_{2x}S_{2x}+S_{2y}S_{2y}}{\mathrm{det}(\mathbf{S})} =& \frac{\vec{S}_2\cdot\vec{S}_2}{\mathrm{det}(\mathbf{S})} \\
\vec{G}^2\cdot\vec{G}^2 &= \frac{S_{1x}S_{1x}+S_{1y}S_{1y}}{\mathrm{det}(\mathbf{S})} =&\frac{\vec{S}_1\cdot\vec{S}_1}{\mathrm{det}(\mathbf{S})} \\
\vec{G}^1\cdot\vec{G}^2 &= -\frac{S_{1x}S_{2x}+S_{1y}S_{2y}}{\mathrm{det}(\mathbf{S})} =&-\frac{\vec{S}_1\cdot\vec{S}_2}{\mathrm{det}(\mathbf{S})}.
\end{aligned}\end{equation}

This result shows that $G^i\otimes G^j$ is the inverse of the Gram matrix $\mathbf{S}$,
\begin{equation}
G^{i}\otimes G^{j} = \frac{1}{\mathrm{det}|\mathbf{S}|}
\begin{pmatrix}
\vec{S}_2\cdot\vec{S}_2 & -\vec{S}_1\cdot\vec{S}_2 \\
-\vec{S}_1\cdot\vec{S}_2 & \vec{S}_1\cdot\vec{S}_1
\end{pmatrix}
=\mathbf{S}^{-1}.
\end{equation}

 We can finally express the 2D right Cauchy-Green tensor (Eq. \ref{Equation:RightCauchyGreen}), needed in section \ref{sec:energy} to calculate the elastic energy,  in terms of the Gram matrices $\mathbf{s}$ and $\mathbf{S}$:
\begin{equation}
\mathbf{C}=\mathbf{F}^T\mathbf{F} = \mathbf{s}\,\mathbf{S}^{-1}.
\label{Equation:RightCauchyGreenExplained}
\end{equation}

We note that Eq. (\ref{Equation:RightCauchyGreenExplained}) can also be used to compute the Green-Lagrange strain tensor $\mathbf{E}=\mathbf{F}^T\mathbf{F}-\mathbf{I}$ from the vertex coordinates. $\mathbf{E}$ converges to the infinitesimal strain tensor $\mathbf{\varepsilon}$ in the limit of small deformations. Eq.\ (\ref{Equation:RightCauchyGreenExplained}) is thus the key result for evaluating strain in  Surface Evolver calculations. We note that Eq.\eqref{Equation:RightCauchyGreenExplained} also gives the correct strain for displacements of vertices normal to the surface.

\subsection{Elastic energy \label{sec:energy}}
In this section we explain how the elastic contribution to the interfacial energy  is determined in our simulations. According to the compressible 3D Neo Hookean model implemented in the Surface Evolver\cite{Bouzidi_CompStruct_2004}, and commonly used in the literature \cite{Pence2015} the elastic energy per volume is
\begin{equation}
W_{3D} = \frac{G}{2} (Tr\, \mathcal{C}-3)-G \ln J +\frac{\Lambda}{2} (\ln J)^2.
\label{eq:3D energy density 1}
\end{equation} 
$G$ and $\Lambda$  are the Lamé parameters.
 $J^2=\mathrm{det}(\mathcal{C})$ is an invariant of $\mathcal{C}$, a scalar quantity independent of the reference frame. It is given by the ratio of the volumes of a material element in the current deformed and initial states.  
In the limit of small deformations, the energy density Eq.\ref{eq:3D energy density 1} reduces as expected to the one deduced from Hooke's law for linear elastic isotropic materials \cite{LandauLifshitz},  using the infinitesimal strain tensor $\mathbf{\epsilon}$ defined by Eq. \ref{eq:epsilon}. 
\begin{equation}
W_{3D} = \frac{ \Lambda}{2} Tr (\,\mathbf{\epsilon})^2+G Tr(\mathbf{\epsilon}^2).
\label{eq:linear_energy_density}
\end{equation}

\par 
The elastic skins considered in our work are so thin that their bending stiffness is negligible. Their resistance to shear deformations where the two opposite faces are displaced relative to each other is very strong, we neglect this mode of deformation and assume a state of {\it plane stress}, consistently with the Kirchhoff hypotheses of thin shell theory \cite{axelrad1987}. Using Cartesian coordinates with an $x_3$ axis perpendicular to an element of the skin, this is expressed as $\mathcal{C}_{31}=\mathcal{C}_{32}=\mathcal{C}_{13}=\mathcal{C}_{23}=0$. 
In the same spirit, we consider the case where the stress normal to the skin has a negligible effect on its shape, so that we can assume $\sigma_{33}=0$ without loss of generality.
For plane stress, the changes of volume and changes of skin thickness are directly related. To analyse this feature, we recall a general relation between the energy density and the Cauchy stress of hyperelastic materials \cite{Mal1991}
\begin{equation}
   J \mathbf{F}^{-1} \mathbf{\sigma}\mathbf{F}^{-T}   =2 \frac{\partial W_{3D}}{\partial\mathcal{C}}.
\end{equation}
The plane stress condition can thus be expressed as 
\begin{equation}
   \frac{\partial W_{3D}}{\partial\mathcal{C}_{33}}=0.
\end{equation}
Using Eq.\ref{eq:3D energy density 1} this yields.  
\begin{equation}
 \Lambda \ln J = G (1 - \mathcal{C}_{33}).
  \label{eq:JC33}
\end{equation}
Physically speaking, this equation previously derived for a similar constitutive equation \cite{Pascon2019} relates the squared ratio of the current and initial skin thicknesses given by $\mathcal{C}_{33}$ to the ratio of the current and initial skin volumes, expressed by $J$. 
In the aim to derive a 2D energy density, we write Eq.\  \eqref{eq:JC33} as a function of the components of $\mathcal{C}$, taking into account that many of them are zero in the case of plane stress, as pointed out above:
\begin{equation}
    \mathcal{C}_{33}(\mathcal{C}_{11}\mathcal{C}_{22}-\mathcal{C}_{12}^2)=\exp\left[\frac{2 G}{\Lambda}(1-\mathcal{C}_{33})\right]
     \label{eq:JC332}
\end{equation}
To represent the skin as a 2D material whose deformation is fully specified by $\mathcal{C}_{11},\mathcal{C}_{22}$ and $\mathcal{C}_{12}$, we need to express $\mathcal{C}_{33}$ in terms of these other variables. This can be done  by solving Eq.\ \eqref{eq:JC332} either numerically \cite{Pascon2019}, or analytically, using Lambert's $W$ function \cite{corless1996}:
\begin{equation}
\mathcal{C}_{33} =   \frac{\Lambda}{2 G }\, W\left[\frac{2 G \, \exp(2 G /\Lambda)}{\Lambda  \mathrm{det}(\mathbf{C}) }\right]\\
= \frac{\Lambda}{2 G }\, W\left[\frac{2 G \, \exp(2 G /\Lambda)}{\Lambda(\mathcal{C}_{11}\,\mathcal{C}_{22}-\mathcal{C}_{12}^2)}\right]
\end{equation}
The latter option has been implemented by R.\ Bouzidi in the Surface Evolver software. Inserting the expression of $\mathcal{C}_{33}$ in Eq.\ \eqref{eq:JC33} and the resulting expression for $\ln J$ into the 3D energy density Eq.\ \eqref{eq:3D energy density 1}, we obtain the following 2D energy density for a neo-Hookean skin, where $h_0$ is the skin thickness in the reference state,
\begin{equation}
W_{2D} =G h_0\left( \frac{1}{2} (Tr\, \mathcal{C}-3)- \frac{G}{\Lambda}(1-\mathcal{C}_{33}) +\frac{G}{2\Lambda} (1-\mathcal{C}_{33})^2 \right).
\label{eq:2D energy density}
\end{equation}
$G h_0$ may be interpreted as a 2D shear modulus.
Neglecting constant terms which are irrelevant for a potential energy and expressing the result in terms of the 2D right Cauchy Green tensor using
$\mathrm{Tr} \mathcal{C} = \mathrm{Tr} \mathbf{C}  + \mathcal{C} _{33}$, we obtain
\begin{equation}
W_{2D} =\frac{G h_0}{2}\left(  Tr\, \mathbf{C} +\mathcal{C}_{33} + \frac{G}{\Lambda} \mathcal{C}_{33}^2 \right).
\label{eq:2D energy density final}
\end{equation}

The skin materials considered in the present paper are much easier to shear than to compress such that $G \ll \Lambda$. In this case, the last term in Eq.\ \eqref{eq:2D energy density final} can be neglected.
 \par Besides the neo-Hookean model discussed so far, the Surface Evolver software provides an alternative energy density expression called "linear elastic model"  which yields  behavior consistent with Eq.\ \eqref{eq:linear_energy_density} in the limit of small deformations. However, one should be aware that for large deformations this numerical model based on the right Cauchy Green tensor  is not consistent with Eq.\ (\ref{eq:linear_energy_density}). 



\section{Pressure-deformation relations of droploons on capillarys expressed via radial stretch}
\label{annex:PressDefNeedle}

In the main body of the article we expressed all relations in terms of area stretch $\lambda_A$. The same approach can be done for the radial stretch $\lambda$ leading, however, to expressions which are less intuitive and less directly accessible by experiments and simulations. For completeness, we shall provide the resulting equations here. 

We can rewrite the interfacial $A$ for a droploon on a capillary larger than a hemisphere as 
\begin{equation}
    \begin{split}
        A & = 2\pi R^2\left(1 - \sqrt{1-\left(\frac{R_n}{R}\right)^2}\right) \\
        & = 2\pi R^2 \mathpzc{f}(R_n/R).
    \end{split}
    \label{eq:interfacialarea}
\end{equation}
The function $F(R_n/R)$ defined by Eq.(\ref{eq:interfacialarea}) helps to express the result in a more concise way.

The term $\ln(A/A_0)$ in the Gibbs relation (\ref{eq:GibbsGamma}) can then be rewritten using Eq. (\ref{eq:interfacialarea}) to give the normalised surface stress of the droploon on the capillary
  \begin{equation}
 \hat{\sigma} = 1+ 2\alpha \ln \lambda + \alpha \ln \xi.
 \label{eq:GibbsNeedle}
 \end{equation}
 The last term, depending on  the geometric factor 
  \begin{equation}
 \xi = \frac{\mathpzc{f}(R_n/R)}{\mathpzc{f}(R_n/R_0)},
 \label{eq:GeomFactor}
 \end{equation}
 expresses the impact of a capillary on the elastic stress at the surface of a sphere, assuming a spherical sector shape.
 
 In the first two terms one recognises the result previously obtained for the perfect sphere (Eq. (\ref{eq:GibbsGamma})).  One can therefore rewrite
   \begin{equation}
 \hat{\sigma} = \hat{\sigma}_{sphere} + \alpha \ln \xi.
 \label{eq:GibbsNeedleSphere}
 \end{equation}
 
Compared to a sphere with the same radius, the presence of the capillary introduces a corrective term in the surface stress which depends on $\alpha$, $R$, $R_n$ and $R_0$. 
 
For neo-Hookean droploons, the droploon shapes on capillaries are no longer perfect spherical sectors, making analytical descriptions much harder - which is why numerical simulations are required. Nevertheless, we shall make here the seemingly crude approximation that the shapes can be approximated as spherical sectors. 

Using exactly the same approach as for the Gibbs interface but with the neo-Hookean relation(see Table \ref{tab:models}), one finds for a neo-Hookean droploon on a capillary 

 \begin{equation}
 \hat{\sigma} = 1 + \frac{\alpha}{3} \left( 1 - \lambda^{-6} \xi^{-3} \right).
 \label{eq:NeoHookeNeedleAnnexe1}
 \end{equation}
 
 After some algebra, this can be rewritten as the expression for the perfect sphere with a corrective term taking account of the capillary
  \begin{equation}
 \hat{\sigma} = \hat{\sigma}_{sphere} + \frac{\alpha}{3} \left(1 - \xi^{-3}\right)\lambda^{-6}.
 \label{eq:NeoHookeNeedleAnnexe2}
 \end{equation}
 
 In the limit of small deformations, our results for both Gibbs and neo-Hooke elastiticy yield the same relation 
  \begin{equation}
 \hat{\sigma} = \hat{\sigma}_{sphere} + \alpha \left(\xi - 1 \right)\lambda,
 \label{eq:HookeNeedleAnnexe3}
 \end{equation}
consistently with what one would obtain for a perfectly spherical sector droploon with Hookean skin on a capillary. In all cases, the corrective term is zero in the reference state where $R=R_0$. Once the interfacial stresses are known, the pressure-deformation relation can be calculated using the Young-Laplace law given in Eq. (\ref{eq:NormalisedPressure}). 

Table \ref{tab:modelsNeedles} summarises normalised expressions derived from this simple geometrical approximation model, together with expressions for the critical stretch.

\begin{table*}[t]
    \centering
    \begin{tabular}{|c|c|c|}
    \hline
   Model on capillary & Normalised surface stress $\hat{\sigma}$ & Critical stretch $\lambda_c$ \\ \hline
Gibbs &  $\hat{\sigma}_{sphere} + \alpha \ln \xi$ & $\frac{R_0\left(1+\sqrt{1-(\frac{R_n}{R_0})^2}\right)e^{-\frac{1}{\alpha}}}{\sqrt{2R_0^2(1+\sqrt{1-(\frac{R_n}{R_0})^2})e^{-\frac{1}{\alpha}}-R_n^2}}$ \\ \hline
Neo-Hooke &  $\hat{\sigma}_{sphere} + \frac{\alpha}{3} \left(1-\xi^{-3}\right)\lambda^{-6}$ & $\frac{R_0(1+\sqrt{1-(\frac{R_n}{R_0})^2})\left(1-\frac{1}{2\alpha} \right)^2}
{ \sqrt{2R_0^2\left(1+\sqrt{1-(\frac{R_n}{R_0})^2}\right)\left(1-\frac{1}{2\alpha} \right)^2-R_n^2}}$ \\ \hline
Hooke &  $\hat{\sigma}_{sphere} + \alpha \left(\xi - 1 \right)\lambda$ & $\frac{R_0(1+\sqrt{1-(\frac{R_n}{R_0})^2})\left(\frac{\alpha}{\alpha+3}\right)^{\frac{1}{3}} }
{ \sqrt{2R_0^2\left(1+\sqrt{1-(\frac{R_n}{R_0})^2}\right)\left(\frac{\alpha}{\alpha+3}\right)^{\frac{1}{3}}-R_n^2}}$\\
\hline
    \end{tabular}
    \caption{Summary of the normalised expressions for the surface stress of drops on capillaries using the approximation that the drop can be described by a spherical sector. While for Gibbs droploons these are correct, they are only approximations for Hookean and neo-Hookean droploons. The expressions for $\hat{\sigma}_{sphere}$ are given in Table \ref{tab:models}. The geometric factor $\xi$ is given in Eq. (\ref{eq:GeomFactor}). }
    \label{tab:modelsNeedles}
\end{table*}

\clearpage



\balance


\bibliography{CapsulePaper.bib} 
\bibliographystyle{rsc} 

\end{document}